\documentstyle[12pt,iopconf,oldlfont]{article}

\def\sigcom{\sigma_{\pi{\scriptscriptstyle N}}}
\def\sigkn{\sigma_{{\scriptscriptstyle KN}}}
\def\MN{M_{{\scriptscriptstyle N}}}
\def\rhob{\rho}
\def\half{{\textstyle{1\over 2}}}
\def\third{{\textstyle{1\over 3}}}
\def\dslash{\partial\llap/}
\def\psibar{\overline\psi}
\def\qbar{\overline q}
\def\mbar{\overline m}
\def\ubar{\overline u}
\def\dbar{\overline d}
\def\sbar{\overline s}
\def\i{{\rm i}}

\begin{document}

\title{Chiral symmetry in nuclei: partial restoration\\
and its consequences}

\author{Michael C. Birse}

\affil{Theoretical Physics Group, Department of Physics and Astronomy,
University of Manchester, Manchester, M13 9PL, UK}

\beginabstract
Partial restoration in nuclear matter of the chiral symmetry of QCD is
discussed together with some of its possible signals. Estimates of corrections
to the leading, linear dependence of the quark condensate are found to be
small, implying a significant reduction of that condensate in matter. The
importance of the pion cloud for the scalar quark density of a single nucleon
indicates a close connection between chiral symmetry restoration and the
attractive two-pion exchange interaction between nucleons. This force is
sufficiently long-ranged that nucleons in nuclear matter will feel a
significant degree of symmetry restoration despite the strong correlations
between them. Expected consequences of this include reductions in hadron
masses and decay constants. Various signals of these effects are discussed,
in particular the enhancement of the axial charge of a nucleon in matter.

\endabstract

\section{Introduction}
Long before the advent of QCD, chiral symmetry was known to be an essential
feature of the strong interaction \cite{ca}. Indeed it was this symmetry
and its associated current algebra that led first to the invention of the
quark model \cite{eight} and thence to QCD itself. Chiral symmetries appear in
theories with massless fermions, where the fields describing right- and
left-handed particles decouple. They are preserved in gauge theories by
interactions with vector fields, at least in the absence of anomalies. In
contrast, interactions with a Lorentz scalar character couple right- and
left-handed fields and so break chiral symmetries.

Our theory of the strong interaction, QCD, possesses an approximate chiral
symmetry because the up and down quarks have current masses, generated by their
coupling to the electroweak Higgs field, that are very much smaller than the
basic energy scale $\Lambda_{QCD}$. The same is true to a lesser extent for the
strange quark. To the extent that we can neglect these masses QCD has separate
isospin symmetries for the right- and left-handed quarks. Hence the symmetry
group is referred to as SU(2)$_R\times$SU(2)$_L$. This can be extended to three
favours although obviously the corresponding symmetry is more strongly broken.

The conservation of the currents associated with these symmetries controls the
form of many of the interactions among pions and nucleons \cite{ca,pagels}. Yet
the symmetry is not obvious in the spectrum of hadrons -- no massless fermions
or parity doublets are seen. Instead the QCD vacuum is not invariant under
chiral rotations and the symmetry is hidden or, to use the standard but
somewhat misleading phrase, ``spontaneously broken" \cite{coleman}. The vacuum
can be thought of as a condensed state of quark-antiquark pairs, with strong
analogies to the condensate of Cooper pairs in a superconductor or the Higgs
vacuum in electroweak theory. The order parameter that describes the hidden
chiral symmetry of the strong interaction is the scalar density of quarks,
often called the quark condensate.

There is an important difference between QCD and a superconductor or Higgs
model: the chiral symmetry is global and its currents are not coupled to gauge
fields. Hence  vacua with different orientations of the order parameter are
distinguishable. For an exact symmetry there would be no restoring force
against chiral rotations of the vacuum and this would lead to the appearance of
massless particles, known as ``Goldstone bosons." In QCD the chiral symmetry is
explicitly broken by the current masses of the quarks and so the corresponding
particles are not exactly massless. Nonetheless the pions have masses that are
very much smaller than all other hadron masses, showing that they are close to
being the Goldstone bosons of hidden chiral symmetry. The kaon masses are
somewhat larger and so those particles are further from being approximate
Goldstone bosons.

As with a superconductor we expect to return to a ``normal" phase where the
symmetry is restored, either at high temperatures or in strong external fields.
The high-temperature phase of QCD is the quark-gluon plasma, which must have
existed in the early universe and which may be recreated in ultra-relativistic
heavy-ion collisions \cite{hwa}. At zero temperature we expect symmetry
restoration when the density of baryons becomes high enough. This might
possibly occur in the cores of neutron stars, converting them to quark stars
\cite{stars}. More importantly, precursors of that transition may already be
present at ordinary nuclear densities. In that case the interior of a nucleus
could be regarded as a laboratory where we can probe the physics of symmetry
restoration in QCD \cite{brown,weise,barev}.

Partial restoration of chiral symmetry inside nuclei could form an important
part of the nuclear binding energy. By reducing the dynamical masses of the
quarks, it could change the masses of nucleons and mesons in the medium and
even their structures. Such modifications would alter the interactions of
nuclei with electromagnetic and weak probes, and would contribute to the
density dependence of nuclear forces.

The starting point for any discussion of chiral symmetry in nuclei is the
expression for the leading density dependence of the the scalar density
of quarks in nuclear matter \cite{dl},
$${\langle \qbar q\rangle_\rho\over \langle \qbar q\rangle_0}=1-{\sigcom\over
f_\pi^2m_\pi^2}\rhob. \eqno(1.1)$$
This form is model-independent \cite{cfg}, but higher-order terms are not. It
expresses the change from the vacuum quark condensate in terms of the baryon
density $\rhob$ and the pion-nucleon sigma commutator $\sigcom$: a measure of
the scalar density of quarks in the nucleon. Taking the recent value
$\sigcom=45\pm 7$ MeV \cite{sigcom} suggests a $\sim 30$\% reduction in the
condensate at nuclear matter densities and a phase transition at about three
times normal nuclear densities. That could have dramatic consequences for the
properties of nucleons and mesons in matter.

Estimates based on the leading density dependence (1.1) should not be taken too
seriously until higher order effects have been calculated. Moreover (1.1)
refers to the spatial average of the quark condensate. The strong short-range
correlations between nucleons could mean that such an average is not a
particularly useful quantity. There are three questions that need to be
addressed:
\begin{itemize}
\item Are there significant corrections to the estimate (1.1) for the quark
condensate inside nuclei?
\item How do correlations between nucleons affect the degree of symmetry
restoration?
\item What are the consequences of partial symmetry restoration for nucleon and
meson properties?
\end{itemize}
None of these has a definitive answer as yet; this review describes the current
state of our understanding and indicates directions for further investigation.

Sec.~2 sets the scene by outlining the basic features of chiral symmetry and
introducing various approaches being used to study its restoration in nuclei.
The quark condensate (Sec.~2.1) and the sigma commutator (Sec.~2.2) are treated
in some detail since, as can be seen from (1.1), they are central to questions
of symmetry restoration in nuclei. The importance of the pion cloud surrounding
a nucleon is stressed since this contributes a long-range component to the
scalar quark density of a nucleon. This provides a connection between symmetry
restoration and the attractive two-pion exchange interaction between nucleons
(Sec.~2.6).

Two models often used in the discussion of chiral symmetry in nuclei are the
linear sigma and the Nambu--Jona-Lasinio models. These are introduced in
Secs.~2.3 and 2.4 respectively. The first of them is based on a scalar,
isoscalar meson field which represents the quark condensate together with pion
fields which provide the corresponding Goldstone bosons. This embodies all the
basic features of hidden chiral symmetry and can be used to illustrate the
qualitative aspects of symmetry restoration. The second attempts to provide a
closer model for QCD by including only quarks, with interactions that
dynamically generate a quark condensate and bind quarks to form mesons.

QCD sum rules have also become popular as a way to relate hadron properties to
condensates, without introducing model assumptions. The basic features of this
approach are described in Sec.~2.5, and illustrated by an application to the
nucleon mass in vacuum. This shows that an important contribution to the
nucleon mass arises from the quark condensate.

Changes of the quark condensate in nuclear matter are described in Sec.~3.1,
starting with the model-independent form (1.1) for the linear density
dependence. Higher-order corrections are examined, first in schematic
treatments based on the linear sigma and NJL models, and then somewhat more
realistically. Although the simple models can provide instructive qualitative
pictures, their quantitative results should not be taken seriously.  More
realistic estimates of  meson-exchange contributions from pions and heavier
mesons suggest that corrections to (1.1) are small, at least at normal nuclear
densities.

The importance of two-pion exchange, which has a fairly long range, means that
correlations between the nucleons should not greatly reduce the effects of
symmetry restoration. In addition it implies a connection between the changes
in the quark condensate and the phenomenological scalar fields of relativistic
nuclear phenomenology. This is also suggested by the application of QCD sum
rules to the self-energy of a nucleon in matter (Sec.~3.4).

The likely effects of partial symmetry restoration on hadrons in matter
(Secs.~3.2 and 3.3) include decreases in hadron masses and meson decay
constants, as well as modifications of nucleon couplings and form factors. The
masses of the approximate Goldstone bosons, pions and kaons, could behave
rather differently. They are of particular interest because of recent
suggestions that $s$-wave kaon condensation could occur in dense nuclear
matter. Present estimates of their behaviour are discussed in Sec.~3.2. For the
nucleon, a decrease in its mass is expected along with changes in the strengths
and form factors for electromagnetic and weak interactions (Sec.~3.3). It has
been suggested that decreases in hadron masses might arise from a universal
scaling related to the scale anomaly of QCD, but in Sec.~3.5 such changes are
shown to be much smaller than those driven directly by the quark condensate.

Possible observable consequences are described in Sec.~4. In many cases, such
as electromagnetic interactions with nuclei, other more conventional mechanisms
also contribute and current calculations are not sufficiently accurate to
distinguish whether symmetry-restoration effects are also present. The
important exception is the axial charge (Sec.~4.1), whose enhancement provides
good evidence for strong scalar fields in nuclei. Other less conclusive signals
surveyed in Sec.~4.2 include quasi-elastic electron scattering and elastic
$K^+$ scattering. Suggestions that the Nolen-Schiffer anomaly, seen in the
energy differences between mirror nuclei, might be a consequence of partial
symmetry restoration are discussed in Sec.~4.3.  Changes in nucleon structure
or meson masses could also have important effects on nuclear forces (Sec.~4.4).
Finally a brief summary is given in Sec.~5.

\section{Chiral symmetry}
In the limit of massless quarks, the QCD Lagrangian is invariant under both
ordinary isospin rotations,
$$\psi\rightarrow(1-\half\i\mbold\beta\cdot\mbold\tau)\psi, \eqno(2.1)$$
and axial isospin rotations,
$$\psi\rightarrow(1-\half\i\mbold\alpha\cdot\mbold\tau\gamma_5)\psi,
\eqno(2.2)$$
where $\mbold{\alpha}$ and $\mbold{\beta}$ denote infinitesimal parameters.
By taking combinations of these rotations involving $1\pm\gamma_5$ we can
independently rotate the isospin of right- and left-handed massless quarks.
Hence the symmetry is referred to as SU(2)$_R\times$SU(2)$_L$. (I concentrate
here on the up and down quarks; the extension to three light flavours is
straight-forward.)

Chiral symmetry is respected by interactions with vector fields (such as gluons
and photons) since $\psibar\gamma_\mu\psi$ is invariant under axial rotations.
The scalar and pseudoscalar densities of quarks are not invariant, transforming
under (2.2) as
$$\psibar\psi\rightarrow\psibar\psi-\mbold\alpha\cdot\psibar\i\mbold\tau
\gamma_5\psi,$$
$$\psibar\i\mbold\tau\gamma_5\psi\rightarrow\psibar\i\mbold\tau\gamma_5\psi+
\mbold\alpha\psibar\psi.\eqno(2.3)$$
Hence fermion mass terms or couplings to scalar fields break the symmetry.

The Noether currents corresponding to the transformations (2.1,2) are the
(vector) isospin current
$${\bf V}^\mu=\psibar\gamma^\mu\half\mbold\tau\psi, \eqno(2.4)$$
and the axial current
$${\bf A}^\mu=\psibar\gamma^\mu\gamma_5\half\mbold\tau\psi. \eqno(2.5)$$
In the absence of current quark masses, these would both be conserved. With
such masses, the divergences of the currents are
$$\partial_\mu V^\mu_i=\half\Delta m\epsilon_{i3j}\psibar\tau_j\psi,
\eqno(2.6)$$
and
$$\partial_\mu A^\mu_i=\mbar\i\psibar\gamma_5\tau_i\psi
+\half\Delta m\delta_{i3}\i\psibar\psi, \eqno(2.7)$$
where $\mbar$ is the average of the current masses for the up and down
quarks, and $\Delta m$ is their difference.

These currents are coupled to photons and W bosons. Hence their matrix elements
can be extracted from the electromagnetic and weak interactions of hadrons. For
example, the weak decay of charged pions involves \cite{pagels}
$$\langle 0|A^\mu_i(x)|\pi_j(q)\rangle=\i f_\pi q^\mu \e^{-\i q\cdot x}
\delta_{ij}, \eqno(2.8)$$
where the pion decay constant is $f_\pi=92.5\pm 0.2$ MeV \cite{fpi,pdg}.

We can imagine a world in which the current masses are zero. Even in that
world, the ``chiral limit," the QCD vacuum would not be invariant under axial
isospin rotations since SU(2)$_R\times$ SU(2)$_L$ is a hidden symmetry. The
pions would then appear as massless Goldstone bosons. In that limit the
Goldstone boson nature of the pions would allow one to determine the
interactions of low-momentum pions purely from chiral symmetry. Another
consequence of the lack of invariance of the vacuum is the non-zero matrix
element (2.8) of the axial current between the vacuum and one-pion states.

The small size of the pion masses compared with those of all other hadrons
indicates that the real world is not too far from the chiral limit. An
essential idea in elucidating the consequences of approximate chiral symmetry
for strong interactions is that of ``partial conservation of the axial current"
(PCAC). An introduction to this can be found in the lectures of Treiman
\cite{treiman} and clear recent discussion of it can be found in
\cite{banerjee}.

PCAC starts from the observation that, with explicit symmetry breaking, the
divergence of Eq.\ (2.8) is
$$\langle 0|\partial_\mu A^\mu_i(x)|\pi_j(q)\rangle=f_\pi m_\pi^2
\e^{-\i q\cdot x} \delta_{ij}. \eqno(2.9)$$
This shows that the operators
$$\mbold\phi(x)=\partial_\mu{\bf A}^\mu(x)/(f_\pi m_\pi^2) \eqno(2.10)$$
connect the vacuum and one-pion states with the same normalisation that
canonical pion fields would have. We can therefore use these operators as
so-called ``interpolating" pion fields. Of course this is a matter of choice:
any operators that can connect these states could be used as interpolating
fields. Note that all such fields should give the same results for all physical
amplitudes involving on-shell pions; where they differ is in their off-shell
extrapolations. The advantage of the PCAC choice is that, by going to the
soft-pion limit $q\rightarrow 0$, we can relate amplitudes involving pions to
the axial transformation properties of the states.

The crucial dynamical assumption embodied in PCAC is that any matrix element of
$\partial_\mu A^\mu_i$ has the form $(q^2-m_\pi^2)^{-1}$ times a smoothly
varying function of $q^2$ \cite{banerjee,treiman}. By assuming that the
variation of these functions over the range $q^2=0$ to $m_\pi^2$ is small, one
can derive various low energy theorems. Corrections to these can be
investigated systematically in increasing powers of the current quark masses, a
technique known as chiral perturbation theory (ChPT)
\cite{pagels,chpt,chptrev}.

\subsection{Quark condensate}
Questions of partial symmetry restoration and its consequences focus on
the changes in the quark condensate in nuclear matter. The starting point is
obviously the quark condensate in the QCD vacuum. A value for this is found by
applying PCAC to the vacuum expectation value of the time-ordered product of
two interpolating pion fields:
$$\third\sum_i\int\!d^4\!x\,\e^{\i q\cdot x}\langle 0|{\rm T}\bigl(
\partial_\mu A^\mu_i(x),\partial_\nu A^\nu_i(0)\bigr)|0\rangle=\i{f_\pi^2
m_\pi^4\over q^2-m_\pi^2}f(q^2),\eqno(2.12)$$
where on the pion mass shell $f(m_\pi^2)=1$. I have taken an isoscalar
combination of fields here so that the result will not be sensitive to the
difference between the quark masses.

The soft-pion limit of this amplitude is obtained by considering pions with
zero three-momentum $\mbold q=0$, and then taking $q^0\rightarrow 0$.
Integrating by parts and taking this limit allows us to rewrite the l.h.s.~of
(2.12) in the form
$$\third\sum_i\langle 0|[Q_5^i,\partial_\nu A^\nu_i(0)]|0\rangle=
\i\third\sum_i\langle 0|\bigl[Q_5^i,[Q_5^i,{\cal H}(0)]\bigr]|0\rangle,
\eqno(2.13)$$
where the $Q_5^i$ are the axial charge operators and ${\cal H}(x)$ is the
Hamiltonian density. If we use the PCAC assumption that
$f(0)\simeq f(m_\pi^2)$ then we find that the soft-pion limit of (2.18) gives
$$\third\sum_i\langle 0|\bigl[Q_5^i,[Q_5^i,{\cal H}(0)]\bigr]|0\rangle
\simeq-f_\pi^2 m_\pi^2. \eqno(2.14)$$
This double commutator picks out the (isoscalar) part of the Hamiltonian that
breaks the symmetry and so (2.14) gives a connection between the pion mass and
the strength of the symmetry breaking, known as a Gell-Mann--Oakes--Renner
(GOR) relation \cite{gmor}. The form of this equation shows that it is an
energy-weighted sum rule; PCAC is equivalent to assuming that it is saturated
by a single state, namely the pion.

The symmetry-breaking part of the QCD Hamiltonian is $\mbar\psibar\psi$. Hence
the GOR relation takes the form
$$\mbar\langle 0|\psibar\psi|0\rangle\simeq-f_\pi^2 m_\pi^2. \eqno(2.15)$$
If we know the quark masses then we deduce a value for the quark condensate.
The current masses of the light quarks have been estimated from hadron mass
splittings and QCD sum rules \cite{qmass}. Unfortunately neither of these
methods is very precise and, in addition, both the masses and the quark
condensate depend on the choice of renormalisation scale. Typical values for
$\mbar$ lie in the range 5--10 MeV, for a scale of 1 GeV. Since $\mbar$ is not
fixed within a factor of two, the quark condensate (usually quoted per quark
flavour) is similarly uncertain:
$$\langle\qbar q\rangle\equiv\half\langle\psibar\psi\rangle
\simeq-(210\  \hbox{MeV})^3\quad\hbox{to}\quad-(260\
\hbox{MeV})^3. \eqno(2.16)$$
The quark condensate in the QCD vacuum is negative; the positive scalar
densities associated with the quarks and antiquarks present in hadronic matter
always tend to cancel some of the condensate, pushing the vacuum towards
symmetry restoration.

As noted by Cohen \etal \cite{cfg}, the Feynman-Hellmann theorem provides a
useful way to think about quark densities in terms of the dependence of
energies on the current quark mass. If $|\Psi(\mbar)\rangle$ is a normalised
eigenstate of the QCD Hamiltonian with energy $E(\mbar)$, then the variational
principle leads to
$$\mbar\langle\Psi(\mbar)|\int\!d^3\!{\bf r}\,\psibar\psi
|\Psi(\mbar)\rangle=\mbar{dE\over d\mbar}
\simeq m_\pi^2{dE\over dm_\pi^2}. \eqno(2.17)$$
The second, approximate equality holds to leading order in $\mbar$ and follows
 from the GOR relation. As an example, consider a zero-momentum pion state:
this has energy $m_\pi$ and so (2.17) gives $\mbar\langle\pi|\int\!d^3\!{\bf r}
\psibar\psi|\pi\rangle\simeq\half m_\pi$. This corresponds to a
(volume-integrated) scalar density for the pion in region of 7--14. This
surprisingly large number shows that the pion is a highly collective state,
rather than a simple $q\qbar$ pair.

\subsection{Sigma commutator}
In studies of medium effects on the quark condensate, a crucial role is played
by the scalar density of quarks inside a nucleon. By analogy with the quark
condensate in previous section, we can define a quantity known as the
pion-nucleon sigma commutator \cite{reya}:
$$\sigcom=\third\sum_i\langle N|\left[Q_5^i,[Q_5^i,H]\right]
|N\rangle,  \eqno(2.18)$$
where $|{\rm N}\rangle$ denotes a zero-momentum nucleon state. The commutator
is equal to
$$\sigcom=\mbar\langle N|\int\!d^3\!{\bf r}\,\psibar\psi|N\rangle.
\eqno(2.19)$$
Hence $\sigcom$ is both the contribution of chiral symmetry breaking to the
nucleon mass and a measure of the scalar density of quarks in the nucleon.

PCAC allows us to relate $\sigcom$ to the soft-pion limit of $\pi N$ scattering
\cite{reya} and hence a value for it can be deduced by extrapolation from
physical $\pi N$ scattering amplitudes. The most recent determination gives
$\sigcom=45\pm7$ MeV \cite{sigcom}, although it should be remembered that there
are inconsistencies between the data sets used in the extrapolation \cite{pin}.
This value is significantly smaller than earlier estimates \cite{koch}, mainly
as a result of a much softer form factor. Gasser \etal \cite{sigcom} find a
radius of about 1.3 fm for the scalar form factor of the nucleon and hence a
roughly 15 MeV difference between $\sigcom$ at $q^2=0$ and the value at the
Cheng-Dashen point, $q^2=2m_\pi^2$.

For $\mbar\simeq 5$--10 MeV, the above value for $\sigcom$ suggests a scalar
density of quarks in a nucleon of about $\langle N|\int\!d^3\!{\bf r}
\psibar\psi|N\rangle\simeq 4$--10. This is at least twice what one would expect
in simple quark models. In relativistic models, such as bag \cite{cbm} or
soliton models \cite{birse}, the valence quark contribution is roughly 2, lower
than the naive result of 3 because the scalar density of relativistic fermions
is reduced by a factor $M/E$ compared to the usual (vector) density.

The enhancement of scalar density deduced from $\sigcom$ over that of the
valence quarks indicates a significant contribution from quark-antiquark pairs
in the nucleon. One can think of the valence quarks distorting the condensate
around them, partially restoring chiral symmetry in their neighbourhood
\cite{clean}. The size and range of this distortion are central to questions of
symmetry restoration in nuclear matter. To estimate its extent, we need to know
the restoring forces acting against symmetry restoration.

Two effects may play important roles in the partial restoration of chiral
symmetry around a nucleon. One is a mean field of scalar, isoscalar mesons
which can be thought of as a direct deformation of the condensate. The other is
the pion cloud of the nucleon. These mechanisms are coupled through the strong
mixing between scalar mesons and two-pion channels.

Many approaches, in particular those based on the linear sigma and NJL models,
treat the meson fields at the classical or tree level, and so omit the pion
cloud of the nucleon. They focus on the role of the scalar, isoscalar meson
which is the chiral partner of the pion. This particle, usually denoted
$\sigma$, is the excitation quantum of the quark condensate. At tree level the
forces acting against symmetry restoration are proportional to the square of
its mass. If this $\sigma$ mass is large, then the strong restoring force leads
to a very small tree-level sigma commutator, as illustrated by the linear sigma
model result in Sec.~2.3. On the other hand, if the sigma mass is low, as in
the NJL model, then the observed sigma commutator can be reproduced without
invoking the meson cloud (Sec.~2.4).

Such low values for the $\sigma$ mass, $M_\sigma\simeq 600$ MeV, are similar to
those used for the phenomenological scalar fields in relativistic models of
nuclei \cite{walecka,relnp}. This invites the identification of those fields
with the change in the quark condensate in matter. However there is no evidence
for such a light scalar in the meson spectrum \cite{meissner}. The most likely
candidate for the chiral scalar particle is the $f_0(1400)$ of the data tables
\cite{pdg} but, as Au, Morgan and Pennington \cite{amp} have pointed out, this
resonance may be better thought of as a broad structure in $\pi\pi$ scattering
at around 1 GeV. The phenomenological scalar fields with much lower masses
should rather be regarded as modelling correlated two-pion exchange, as
discussed in Sec.~2.6.

A large mass of 1 GeV or more for the chiral $\sigma$ field suggests a strong
force acting against chiral symmetry restoration. If the coupling to pions were
neglected then any changes in the quark condensate would be small and very
short-ranged. In any chirally symmetric model of nucleon structure, the pion
cloud contributes significantly to many observables, and in particular to
$\sigcom$, because of the large scalar density in the pion mentioned in the
previous section. These contributions are long-ranged because the pions are
light. Calculations of $\sigcom$ in cloudy bag \cite{jameson} and
nontopological soliton models \cite{bmcg} (see also the linear sigma model
calculation discussed in Sec.~2.6) find an extra 20--25 MeV from the
cloud which, combined with the valence part, gives agreement with the observed
value.  Empirical support for a substantial pion-cloud contribution comes from
the large radius for the scalar density of the nucleon found in recent analyses
\cite{sigcom,phs}.

Calculations of $\sigcom$ from first principles using lattice QCD are still at
a preliminary stage. Older calculations, using the quenched approximation and
with rather large current masses, found values similar to the valence
contribution only \cite{latsig1}. More recent results for dynamical fermions
show significant disconnected contributions to $\sigcom$ \cite{latsig2} which
suggest that meson loop effects are important even at the current quark masses
of about 35 MeV that can be reached in present lattice calculations. Note that
the pion cloud contains pieces corresponding to connected diagrams (the only
ones present in the quenched approximation) as well as disconnected ones
\cite{quench,bmcg}. Hence one cannot simply interpret these diagrams as valence
and sea contributions respectively.

Another indicator of the importance of the cloud is the long-standing
discrepancy between estimates of $\sigcom$ from the spectrum of octet
baryons and those from $\pi N$ scattering. If one assumes that the octet
splittings are given by first-order perturbation theory in the current
masses and that there are no strange quarks in the proton, then the
$\sigma$ commutator is \cite{cheng}
$$\sigma_0={\mbar\over m_s-\mbar}\left(M_{\scriptscriptstyle \Xi}
+M_{\scriptscriptstyle \Sigma}-2\MN\right). \eqno(2.20)$$
More generally, one can allow for a nonzero density of strange quarks
and write
$$\sigcom={\sigma_0\over 1-y}, \eqno(2.21)$$
where
$$y={\langle N|\sbar s|N\rangle\over\langle N|\ubar u
+\dbar d|N\rangle}. \eqno(2.22)$$
If one takes $m_s/\mbar=25$ \cite{qmass}, then the observed baryon masses lead
to $\sigma_0\simeq 25$ MeV. Naively this would suggest a large strange-quark
content in the proton, $y\sim 0.5$, and hence a huge contribution to the
nucleon mass from strange quarks \cite{dn}.  Before taking such a conclusion
seriously, one should examine whether first-order perturbation theory is valid
for the baryon masses. As first pointed out by Jaffe \cite{jaffe} in a chiral
bag model and subsequently shown in other models with strong meson clouds
\cite{strq}, there are strong nonlinearities in the dependence on the current
masses. Kaon-cloud effects on baryon observables are very much smaller than
those of pions. Hence the estimate (2.20), which is dominated by the
strange-quark density of the hyperons, gives essentially the valence-quark
piece of $\sigcom$ only.

Gasser \cite{gasser,qmass} has estimated corrections to (2.20) using ChPT.
Non-analytic dependences of baryon masses on $\mbar$ (or equivalently
$m_\pi^2$) arise from the longest-range parts of the pion cloud. Chiral
symmetry requires that logarithmic terms appear only at the order
$m_\pi^4\ln m_\pi$. The leading non-analytic term in $\sigcom$ is of
order $m_\pi^3$:
$$\sigcom=Cm_\pi^2-{9\over 64\pi^2}{m_\pi^3\over f_\pi^2}+\cdots. \eqno(2.23)$$
Gasser's results indicate that such terms can raise the estimate to
$\sigma_0\sim 35$ MeV. Moreover that calculation includes $\pi N$ loops only.
Virtual $\Delta\pi$ states have long be known to give significant contributions
to nucleon properties in the cloudy bag \cite{cbm}, and recently Jenkins and
Manohar have pointed out their importance in ChPT \cite{jenman} (see also:
\cite{jm2,cohbr,bkm}). In the case of $\sigcom$, bag and soliton models give
$\Delta\pi$ contributions of 6--10 MeV \cite{jameson,bmcg}. These can remove
the
remaining discrepancy between the two estimates of $\sigcom$ without requiring
any substantial strange-quark content in the nucleon.

A similar kaon-nucleon sigma commutator can be defined as
$$\sigkn=\half(m_u+m_s)\langle N|\int\!d^3\!{\bf r}\,(\ubar u+\sbar s)
|N\rangle,\eqno(2.24)$$
where an average over proton and neutron is understood. In principle this
could be extracted from KN scattering, but the much larger extrapolations
involved mean that $\sigkn$ is very poorly determined \cite{sigkn}. If the
strange-quark content in the proton is small, $\sigkn$ is to a good
approximation just $m_s/4\mbar$ times $\sigcom$. This estimate suggests that
$\sigkn$ is at least $\sim 280$ MeV.

\subsection{Linear sigma model}
Although lattice QCD can give us essential information about the chiral phase
transition at high temperatures \cite{ftlqcd}, high density calculations are
still in a very preliminary state \cite{barbour} since the Monte-Carlo
integration methods have difficulty coping with chemical potentials. We are
therefore forced to use QCD-motivated models to study that regime. The simplest
model for the physics of symmetry restoration is the linear sigma model
\cite{gml}, which has a venerable history of applications to the hidden chiral
symmetry of the strong interaction.

The model is based on a scalar, isoscalar field that represents the quark
condensate, together with pseudoscalar, isovector pion fields. These transform
like the corresponding quark bilinears (2.3) under axial rotations:
$$\sigma\rightarrow\sigma-\mbold\alpha\cdot\mbold\phi,\qquad\quad
\mbold\phi\rightarrow\mbold\phi+\mbold\alpha\sigma.        \eqno(2.25)$$

These fields can be coupled in a chirally invariant way to an isospin doublet
of
fermions. The fermions can be either nucleons, if we are interested in nuclear
physics, or quarks, if we want to describe baryon structure \cite{birse}.
Here I use the original version involving nucleon fields \cite{gml}. In either
case the model Lagrangian takes the form
$${\cal L}=\psibar\bigl[\i\dslash+g(\sigma+i\mbold\phi\cdot
\mbold\tau\gamma_5)\bigr]\psi+\half(\partial_\mu\sigma)^2+\half(\partial_\mu
\mbold\phi)^2-U(\sigma,\mbold\phi). \eqno(2.26)$$
The symmetry is hidden if the potential $U$ is chosen to have a ``Mexican-hat"
form:
$$U_0(\sigma,\mbold\phi)={\lambda^2\over 4}\big(\sigma^2+\mbold\phi^2-\nu^2
\big)^2.\eqno(2.27)$$
As can be seen from Fig.~2.1, there is a circle of degenerate minima in the
brim of the hat. If we take the physical vacuum to have good parity
($\sigma=\pm \nu$, $\mbold\phi=0$) there is no restoring force against pionic
excitations about the vacuum. The pions are thus massless Goldstone bosons. The
$\sigma$ field experiences a strong restoring force and so its excitations
are massive scalar mesons.

We can give the pions their observed masses by tipping the Mexican hat so that
the symmetry is broken and the vacuum is unique. The full potential is then
$$U(\sigma,\mbold\phi)={\lambda^2\over 4}\big(\sigma^2+\mbold\phi^2-\nu^2
\big)^2+f_\pi m_\pi^2\sigma.   \eqno(2.28)$$
With this choice of symmetry-breaking term, the model explicitly embodies
PCAC: the divergence of the axial current is proportional to the pion field
of the Lagrangian.

The matrix element for pion decay in the model fixes the vacuum expectation
value of the $\sigma$ field to be $-f_\pi$. The nucleon mass is thus
$$\MN=gf_\pi, \eqno(2.29)$$
where $g$ is the $\pi N$ coupling constant. This is just the
Goldberger--Treiman
relation with $g_A=1$. In terms of the parameters of the potential, the meson
masses are
$$m_\sigma^2=\lambda^2(3f_\pi^2-\nu^2),\qquad\qquad
m_\pi^2=\lambda^2(f_\pi^2-\nu^2). \eqno(2.30)$$

To illustrate of some of the features of this model, I revisit the calculation
of soft-pion scattering from a nucleon in this model. The results of this will
be used in Sec.~3. The diagrams that contribute at tree-level are shown
in Fig.~2.2. Consider, for simplicity, scattering of a virtual pion of zero
three-momentum but with energy $\omega$ from a nucleon at rest. Using (2.29,
30), the scattering amplitude can be written
$$T={\MN\over f_\pi^2}\left[{m_\sigma^2-m_\pi^2\over m_\sigma^2}
-{4\MN^2\over 4\MN^2-\omega^2}\right]. \eqno(2.31)$$
Chiral symmetry ensures that terms of order $m_\pi^0$ (and $\omega^0$) cancel.
For soft pions, with $\omega=0$, this amplitude takes the form required by PCAC
\cite{ca,reya}:
$$T=-{\sigcom\over f_\pi^2}, \eqno(2.32)$$
where, at tree-level, the sigma commutator is
$$\sigcom={m_\pi^2\over m_\sigma^2}\MN. \eqno(2.33)$$

This result (2.33) for $\sigcom$ can of course also be obtained directly
as the matrix element of the symmetry-breaking term $f_\pi m_\pi^2\sigma$.
For a typical $\sigma$ mass of 1200 MeV, it gives $\sigcom\simeq 13$ MeV.
This is similar to the valence-quark contributions discussed in the previous
section, and it indicates the need to go beyond tree-level by including
loop diagrams corresponding to the pion cloud.

In the soft-pion limit, the contributions of diagrams 2.2(b) and (c) come from
negative-energy intermediate states. In time-ordered perturbation theory such
diagrams can be interpreted in terms of virtual nucleon-antinucleon states and
hence the cancellation between these Z-graphs and $\sigma$ exchange is often
referred to as ``pair suppression."

Both scalar fields and Z-graphs play important roles in relativistic treatments
of nuclei \cite{walecka,relnp,dirph,ray}. It should be remembered that they
describe short-distance physics which is not determined purely by chiral
symmetry and which is thus model-dependent. The $\sigma$ field is needed if one
wants to describe chiral symmetry restoration using an effective field theory
of nucleons and mesons. The Z-graphs are essential if current conservation and
the associated low-energy theorems are to be maintained. Of course nucleons are
not point-like Dirac particles and so it is likely that the interpretation of
these diagrams as pair creation should not be taken too literally. Brodsky
\cite{brod} has long argued that form factors should suppress pair creation of
composite objects. Instead the Z-graphs should be regarded as mocking up the
effects of nucleon structure \cite{zgraph}. Calculations in a nontopological
soliton model \cite{let} have shown that a combination of quark Z-graphs and
excitations leads to the same results for several low-energy theorems.

The scattering amplitude (2.32) for soft pions is repulsive. From (2.22) one
can see that at tree-level this repulsion increases with the pion energy. This
increase arises from the energy denominators of the Z-graphs. In contrast the
amplitude at the physical pion threshold (zero pion three-momentum and
$\omega=m_\pi$) is very small \cite{pin}. Such a reduction can be obtained by
including positive-energy intermediate states: pion loops or excited
baryons are needed for a realistic description of $\pi N$ scattering.

\subsection{Nambu--Jona-Lasinio model}
The sigma model can give us a good qualitative picture of the physics involved,
but in QCD we have no fundamental scalar fields; really the $\sigma$ and pion
fields should be regarded as approximate descriptions of the corresponding
bilinear combinations of quark fields. A model that is often proposed as a
step closer to QCD is the remarkable one introduced by Nambu and Jona-Lasinio
\cite{njl} in which chiral symmetry is is hidden dynamically. It consists of
fermions interacting via a local four-fermion  interaction (a zero-range
two-body force):
$${\cal L}_{\rm NJL}=\psibar(\i\dslash-\mbar)\psi
+{G\over 2}\left[(\psibar\psi)^2+
(\psibar\i\mbold\tau\gamma_5\psi)^2\right].  \eqno(2.34)$$
In the modern versions of this the fermions are interpreted as quarks
\cite{njlrev}. The combination of scalar and pseudoscalar interactions is
chosen since it is chirally symmetric. Obviously such a zero-range force is a
caricature of the strong gluon-exchange interactions between quarks.
Nonetheless it retains some of their important features, with the notable
exception of confinement. Since it is non-renormalisable, the model only makes
sense with a cut-off at short distances. A wide variety of cut-off procedures
has been used \cite{mrag}; fortunately the qualitative results do not depend on
this choice. The model has also been extended to include strange quarks as
well as vector and axial-vector interactions between the quarks \cite{njlrev}.

In the one-loop approximation (a Hartree treatment of the Dirac sea), the
dynamical quark mass satisfies the nonlinear equation,
$$M_q=\mbar+4N_cN_fG\int^\Lambda\!{d^4\!k\over(2\pi)^4}{M_q\over k^2+M_q^2},
\eqno(2.35)$$
for $N_c$ colours and $N_f$ flavours of light quarks. I have given the form for
a simple covariant cut-off, $k\le\Lambda$, on the magnitude of the
four-momentum (which has been Wick-rotated to Euclidean space-time). This
equation always has the obvious solution $M_q=0$, corresponding to a vacuum
with manifest chiral symmetry. For values $G$ greater than a critical
$G_c(\Lambda)$, which depends on the cut-off, the lowest-energy solution gives
a vacuum in which the quarks have a non-zero mass.

In the latter vacuum, the chiral symmetry is hidden. There is a quark
condensate,
$$\langle\qbar q\rangle\equiv{1\over N_f}\langle\psibar\psi\rangle
=-4N_c\int^\Lambda\!{d^4\!k\over(2\pi)^4}{M_q\over k^2+M_q^2}, \eqno(2.36)$$
whose coupling to the quarks produces their mass (2.35), analogously
to the vacuum expectation value of $\sigma$ in the linear sigma model. The
vacuum of the NJL model is a condensate of quark-antiquark pairs, and the
energy difference $2M_q$ between the top of the Dirac sea and the lowest
valence quark level can be thought of as the the gap energy required to break a
pair and form a quasiparticle-hole excitation.

The local nature of the interaction in this model means that the Bethe-Salpeter
equation for quark-antiquark scattering has a simple form and can be solved by
summing a geometric series. The bound-state poles of this amplitude can be used
to determine meson masses and wave-functions. With no current quark masses the
pions are massless Goldstone bosons. and the $\sigma$ meson is an almost
unbound
state in the scalar, isoscalar channel, with a mass
$$m_\sigma\simeq 2M_q. \eqno(2.37)$$
Current quark mass terms can be added to the Lagrangian (2.34), breaking the
symmetry and giving the pions masses.

The model parameters are chosen to fit the pion decay constant and mass and
give a dynamical quark mass in the range 300--400 MeV. The corresponding values
for the cut-off are about 600--800 MeV, depending on the form used. These give
reasonable values for the quark condensate, between $-$(220~MeV)$^3$ and
$-$(290~MeV)$^3$.

Baryons can be constructed as solitons in this model \cite{njlsol}. This is
generally done by bosonising the model, converting it into an equivalent model
involving only meson fields. Auxiliary $\sigma$ and pion fields are introduced
to express the Lagrangian in a form that is bilinear in the quark fields,
$${\cal L}'_{\rm NJL}=\psibar\bigl[\i\dslash+g(\sigma+i\mbold\phi\cdot
\mbold\tau\gamma_5)\bigr]\psi-\half\mu^2(\sigma^2+\mbold\phi^2). \eqno(2.38)$$
Integrating out the quark fields then leaves a purely bosonic effective
action,
$$S_{\rm NJL}=-\i{\rm Tr}\ln\bigl[\i\dslash+g(\sigma+i\mbold\phi\cdot
\mbold\tau\gamma_5)\bigr]-\half\mu^2\int\!d^4\!x\,(\sigma^2+\mbold\phi^2).
\eqno(2.39)$$
The first term is the logarithm of the determinant of the Dirac operator. It is
a complicated object whose dependence on the boson fields is highly nonlocal
because of the effects of vacuum polarisation. Techniques have been developed
for evaluating this determinant for localised soliton configurations and then
minimising the effective action \cite{njlsol}. For uniform systems things are
much simpler: the nonlocal terms do not contribute and effective action reduces
to an effective potential. If the coupling strength is greater than the
critical value, this potential has a similar form to the Mexican hat of the
linear sigma model (although it is not simply a quartic function of the
fields).

An obvious shortcoming of this model is that it does not absolutely
confine quarks. Hence, for example, mesons with masses greater than
$2M_q$ are not bound. Another problem is the lightness of the $\sigma$ meson,
typically around 700 MeV. Although it may be tempting to identify this with the
light $\sigma$ of nuclear physics, that should be avoided for the reasons
discussed in Secs.~2.2 and 2.6. The low $\sigma$ mass means that the NJL model
underestimates the forces reacting against symmetry restoration. This results
in large vacuum polarisation effects, even without inclusion of the pion cloud.
For example, the pion-quark sigma commutator is about twice the naive
expectation: $$\sigma_{\pi q}=\mbar{dM_q\over d\mbar}\simeq 2\mbar.
\eqno(2.40)$$ Also, the dressed quark has a scalar form factor whose radius is
determined by the $\sigma$ mass and is therefore large \cite{klvw}.

The softness of the vacuum in the NJL model can be removed by adding extra
interactions. For example, Ripka and Jaminon \cite{ripjam} have suggested a
variant in which the quadratic term in the ``half-bosonised" Lagrangian (2.38)
is replaced by a quartic. The model was originally motivated by ideas of scale
invariance, but in fact the extra dilaton field plays very little role in the
dynamics see Sec.~3.4 below). The quartic term could be thought of as arising
 from a four-body interaction among the quarks. It is equivalent to adding an
extra Mexican-hat term to the effective potential. This has the effect of
increasing the $\sigma$ mass to $\sim 1.5$ GeV and correspondingly reducing the
vacuum polarisation in this channel.

\subsection{QCD sum rules}
A rather different approach from the models just described is provided by the
QCD sum rules developed by Shifman, Vainshtein and Zakharov \cite{svz}. These
attempt to relate hadron properties to condensates: vacuum expectation values
that describe the non-perturbative aspects of the QCD vacuum. Recently these
sum rules have become popular as a possible tool for studying the behaviour of
hadrons in nuclear matter.

The GOR relation of Sec.~2.1 can be thought of as a prototype for these sum
rules: it uses a Green's function of interpolating fields to relate pion
properties and the quark condensate. It is particularly simple because chiral
symmetry ensures that only one state (the pion) dominates the propagator at low
$q^2$ and that only one condensate appears. For interpolating fields
corresponding to other mesons or baryons, such simplifications do not occur.

Consider a general Green's function of the form
$$\Pi(q)=\i\int\!d^4\!x\,\e^{\i q\cdot x}\langle 0|T(\eta(x),\overline\eta(x))
|0\rangle, \eqno(2.41)$$
where $\eta(x)$ denotes an interpolating field with the quantum numbers of the
hadron of interest. By inserting a complete set of states, one can express this
in the form of a dispersion relation involving the spectral density of states
in the chosen channel. This density can then be written in terms of the masses
and couplings of the hadrons. Alternatively the operator-product expansion
(OPE) \cite{ope} can be used to express the Green's function as a sum of vacuum
matrix elements of local operators. These operators are combinations of quark
and gluon fields whose matrix elements are the condensates representing the
nonperturbative physics. Each is multiplied by a function of $q^2$ which can be
calculated from perturbative QCD. By matching the two expressions for the same
propagator, values for the condensates can be deduced from observed hadron
properties.

The OPE is valid for large space-like momenta ($q^2<0$), but in that region
many resonances contribute to the spectral representation of the propagator. A
direct comparison of the two forms as functions of $q^2$ is not practical;
instead a weighted integral is used \cite{rry,qssr}. The art of the sum rule
approach is to pick a weighting function that both emphasises the role of
low-lying resonances in the spectral representation and keeps down the number
of condensates making important contributions to the OPE. The choice of Shifman
\etal\cite{svz}, which has been found to be particularly convenient, is the
Borel transform. Although this can be expressed as a contour integral
\cite{cfgsr}, it is normally written as
$$ {\cal B}f(Q^2)=\lim_{Q^2,n\rightarrow\infty}\left(-{d\over dQ^2}\right)^n
f(Q^2)\equiv \hat f(M^2), \eqno(2.42)$$
where $Q^2=-q^2$ and $n$ are taken to infinity while their ratio
$$M^2={Q^2\over n}, \eqno(2.43)$$
is kept constant. The result obviously depends on an arbitrary parameter $M$,
the ``Borel mass."

This transform eliminates any polynomials in $Q^2$ which arise from
subtractions in the dispersion relation. More importantly, it exponentially
suppresses contributions from high-mass states in the spectral representation.
In the OPE, high dimension condensates are suppressed by inverse powers of
$M^2$. While, on one hand one would like $M$ to be small enough that the lowest
state dominates in the spectral representation, one would also like $M$ to be
large to keep the number of condensates in the OPE manageable. The exponential
suppression of higher resonances suggests that one may be able to find an
intermediate range for $M$ where these conflicting desires can be reasonably
well satisfied. The fact the results should be independent of $M$ provides a
consistency check: if there is a region where the two representations agree and
are flat functions of $M$, then one may have some confidence in the deduced
values  for the condensates.

To illustrate the sum rule concept, I outline here their application to the
nucleon. Detailed discussions can be found in the review \cite{rry} and in
Ref.~\cite{cfgsr}. There are two linearly independent ways is which an
interpolating nucleon field can be constructed from one down- and two up-quark
fields coupled to spin-$\half$ and isospin-$\half$. As with the Borel mass,
one tries to find a field that both ensures that the nucleon dominates in the
spectral representation and minimises the higher-order contributions in the
OPE. The optimal choice is the one introduced by Ioffe \cite{ioffe} which can
conveniently be expressed in a form where the up quarks are coupled to form a
vector diquark:
$$\eta(x)=\epsilon_{abc}[u^T_a(x)C\gamma_\mu u_b(x)]\gamma_5\gamma^\mu d_c(x),
\eqno(2.44)$$
where $a$, $b$, $c$ label the colours of the quark fields and $C$ is the charge
conjugation matrix. From Lorentz covariance, parity and time-reversal
symmetries, the propagator (2.41) constructed with this field must have the
form
$$\Pi(q)=\Pi_s(q^2)+{q\llap/}\Pi_q(q^2). \eqno(2.45)$$
Using a dispersion relation, each of the scalar functions $\Pi_i(q^2)$ can be
written as a nucleon pole plus continuum:
$$\Pi_s(q^2)={\lambda_N^2\MN\over q^2-\MN^2}-\int_{s_0}^\infty
ds\,{\rho^{\rm cont}_s(s)\over q^2-s}+\cdots,  \eqno(2.46{\rm a})$$
$$\Pi_q(q^2)={\lambda_N^2\over q^2-\MN^2}-\int_{s_0}^\infty
ds\,{\rho^{\rm cont}_q(s)\over q^2-s}+\cdots,  \eqno(2.46{\rm b})$$
where the dots denote polynomials from subtractions.
Alternatively the OPE yields the following for each of the terms
$$\Pi_s(q^2)={1\over 4\pi^2}q^2\ln(-q^2)\langle\qbar q\rangle+\cdots,
\eqno(2.47{\rm a})$$
$$\Pi_q(q^2)=-{1\over 64\pi^2}q^4\ln(-q^2)-{1\over 32\pi^2}\ln(-q^2)
\langle{\alpha_s\over\pi}G_{\mu\nu}G^{\mu\nu}\rangle+\cdots.
\eqno(2.47{\rm b})$$
where I have shown explicitly only the contributions from the quark and gluon
condensates with dimension less than five. Higher-dimensioned condensates
appear multiplied by inverse powers of $q^2$. Their full forms can be found in
\cite{iofsmi,cfgsr}.

The pole term in the spectral representation introduces an unknown strength
$\lambda_N$ for the coupling of the interpolating field to the nucleon.
The continuum in the spectral representation (2.46) is often approximated by
the perturbative continuum, arising from the cuts in the OPE expressions,
starting at $q^2=s_0$ which is treated as an adjustable parameter. Obviously
this is not a particularly accurate representation, but provided the Borel mass
is not too large the weighted average in the sum rule is not very sensitive to
the details of the continuum.

Equating the two expressions (2.46, 47) for each of the terms in the propagator
and Borel transforming leads to two sum rules:
$$\lambda_N^2\MN\e^{-\MN^2/M^2}+\int_{s_0}^\infty ds\,\e^{-s/M^2}
\rho^{\rm cont}_s(s)=-{M^4\over 4\pi^2}\langle\qbar q\rangle+\cdots,
\eqno(2.48{\rm a})$$
$$\lambda_N^2\e^{-\MN^2/M^2}+\int_{s_0}^\infty ds\,\e^{-s/M^2}
\rho^{\rm cont}_q(s)={M^6\over 32\pi^4}+{M^2\over 32\pi^2}
\langle{\alpha_s\over\pi}G_{\mu\nu}G^{\mu\nu}\rangle+\cdots.
\eqno(2.48{\rm b})$$
The continuum contribution can be taken over to the r.h.s.~of each equation,
where it modifies the coefficients of the lowest condensates. The
renormalisation group can be used to improve these sum rules by summing up the
leading logarithms of $Q^2$ in the coefficients of the condensates.

Ioffe's sum rule \cite{ioffe} for the nucleon mass is given by the
ratio of these two sum rules in which the coupling to the interpolating field
cancels. An oversimplified but instructive version of this \cite{rry} is
obtained by dropping all but the quark condensate in the OPE and neglecting
the continuum. This leaves an expression for the nucleon mass that is
proportional to the quark condensate:
$$\MN=-{8\pi^2\over M^2}\langle \qbar q\rangle. \eqno(2.49)$$
Taking a typical value of 1 GeV for the Borel mass and the quark condensate
 from (2.16), one finds a nucleon mass in the region of 900 MeV. This shows
that
the quark condensate makes a major contribution to the nucleon mass, in
agreement with expectations based on the chiral models of the previous
sections. Since the sum rule (2.49) obviously depends strongly on $M$, higher
condensates are essential if it is to yield quantitative results.

In fact there are strong cancellations among the terms neglected in (2.49),
which is why it gives a reasonable estimate of the nucleon mass. The most
important piece omitted from it is the four-quark condensate, $\langle (\qbar
q)^2\rangle$, which will be needed for the discussion of sum rules in matter.
Like other four-quark and higher condensates, this is often estimated
using the factorised or ``vacuum dominance" ansatz \cite{svz},
$$\langle (\qbar q)^2\rangle=\langle\qbar q\rangle^2. \eqno(2.50)$$
However, even in the vacuum, there are indications that this assumption is
violated by about a factor of two \cite{qssr}. We shall see that the
uncertainties associated with this condensate place severe limitations on the
predictive power of QCD sum rules for hadrons in matter.

One should also remember that QCD sum rules focus on short-distance physics, as
described by a truncated OPE. Long-range physics is subsumed into a simple
parametrisation of the spectral function and so is not treated reliably. As
pointed out by Griegel and Cohen \cite{gc}, the simple ansatz for the continuum
described above means that pion-cloud contribution to the nucleon mass is not
properly described. In particular, the nonanalytic dependence of the nucleon
mass on the current quark masses \cite{gasser,qmass} is not reproduced. The
resulting uncertainty of $\sim 100$ MeV may not be too serious for the nucleon
mass, but it does mean that the sigma commutator cannot be reliably estimated
 from QCD sum rules.

\subsection{NN interaction}

As alluded to in previous sections, scalar fields play a central role in
relativistic nuclear physics. These are scalar, isoscalar fields, usually
denoted by $\sigma$, that provide the attractive forces between nucleons in
relativistic models of nuclear structure \cite{walecka,relnp} or nucleon
scattering from nuclei \cite{dirph,ray}. Their similarity to the fields used to
model the quark condensate invites the question of whether such fields are
related to partial restoration of chiral symmetry. Such a relationship has been
suggested in the context of QCD sum rules \cite{cfgsr}, the linear sigma model
\cite{cs1} and the Nambu--Jona-Lasinio (NJL) model \cite{cs2}.

Phenomenological $NN$ potentials \cite{nn,paris,bonn} also find substantial
intermediate range attraction in the scalar isoscalar channel. This
attraction has long been known to be well described by exchange of two
correlated pions \cite{nn,djv,lin}. The light $\sigma$ particle of relativistic
phenomenologies should be regarded as modelling this two-pion exchange process,
and not an indication of a light chiral partner for the pion. Furthermore
models with a light chiral partner for the pion give rise to strong many-body
forces between nucleons \cite{nyman} and cannot provide an acceptable
description of nuclear properties \cite{fs}.

Although the scalar fields of relativistic nuclear models are introduced to
describe very different physics from the chiral partner of the pion, there is
strong mixing of the chiral scalar with two-pion states, and this sharp
distinction is lost. To explore the relationship between the attractive $NN$
force and chiral symmetry restoration, I look at the calculation of two-pion
exchange in the linear sigma model \cite{lin,cps,btwo}. Other relevant work on
chiral symmetry and the two-pion exchange interaction can be found in
Ref.~\cite{nnchi}.

The interaction of interest can be found from the scalar, isoscalar piece of
the irreducible scattering amplitude for two nucleons. The simplest
contribution to this is just direct $\sigma$-exchange, Fig.~2.3. At one-loop
order there are four diagrams involving exchange of a pair of virtual
pions between the nucleons \cite{btwo}. These are shown in Fig.~2.4. Working to
this order, I have not included interactions between the exchanged pions. Such
interactions are known to be essential to the strong attraction between the
nucleons \cite{nn,djv} and hence this calculation cannot yield a realistic
result for the full scalar interaction.

Direct $\sigma$ exchange, Fig.~2.3, is purely scalar and isoscalar. Its
contribution to the $NN$ $T$-matrix is easily evaluated giving
$$S_D=-\left({\MN\over f_\pi}\right)^2 {1\over m_\sigma^2-t}. \eqno(2.51)$$
The evaluation of the loop diagrams is long and tedious; details can be found
in Ref.~\cite{cps}. The scalar, isoscalar pieces of the $T$-matrix
corresponding to Figs.~2.4(a)-(d) are denoted here by $S_L$, $S_V$, $S_X$ and
$S_B$ respectively.

The amplitude for the box diagram, Fig.~2.4(d), arises from iterating one-pion
exchange in the Bethe-Salpeter equation. To get an irreducible amplitude, the
iterated one-pion exchange must be removed. However one cannot simply drop the
whole contribution of 2.4(d) since, with pseudoscalar (PS) $\pi N$ coupling;
that would leave an irreducible amplitude that would not satisfy ``pair
suppression" (the constraints imposed by chiral symmetry mentioned in
Sec.~2.3). That would not matter if the scattering equation were treated
exactly, but any approximation would produce large violations of chiral
symmetry. To avoid such problems phenomenological treatments of the $NN$
interaction normally use pseudovector (PV) $\pi N$ coupling
\cite{tw,horo,bonn}. Subtracting the box diagram calculated with PV coupling
leaves an irreducible scalar amplitude that can be compared with such
interactions \cite{lin}.

The full irreducible amplitude can be written
$$S=S_D+S_L+2S_V+S_X+S_B-S_{B(PV)}, \eqno(2.52)$$
where the detailed forms of the individual amplitudes can be found in
Ref.~\cite{btwo}. Although the contributions from individual diagrams are
large, there are strong cancellations between them as required by chiral
symmetry. The net strength is about half that of a typical phenomenological
$\sigma$ exchange amplitude. This just shows that shows that interactions
between the exchanged pions ought to be included, together with diagrams
involving intermediate $\Delta$'s \cite{djv,bonn}.

Of most interest for the present discussion is the piece involving direct
$\sigma$ coupling to one of the nucleons. These describe the degree of
symmetry restoration experienced by that nucleon and is given by the
sum of three diagrams:
$$S_{CR}=S_D+S_L+S_V. \eqno(2.53)$$
The strength of this at zero momentum transfer is related to the sigma
commutator, since in this model $\sigma_{\pi{\scriptscriptstyle {\rm N}}}$ is
just proportional to the matrix element of the $\sigma$ field in a nucleon.
Specifically one has
$$S_{CR}(t=0)=-{\MN\over f_\pi^2 m_\pi^2}\sigma_{\pi{\scriptscriptstyle {\rm
N}}}. \eqno(2.54)$$
The contribution from the pionic diagrams correspond to the cloud contributions
to $\sigma_{\pi{\scriptscriptstyle {\rm N}}}$ calculated in chiral bag and
soliton models \cite{jameson,bmcg}, and for a cut-off of $\Lambda\simeq 1$ GeV
they have a similar magnitude. They more than double $\sigcom$ compared with
the tree-level result (2.33) which on its own gives $S_D$.

The form of this result (2.54) shows that it is much more general than the
model studied here. It follows from the assumption that the nucleon mass is
proportional to the quark condensate, and the fact that $\sigcom$ is a measure
of the scalar quark density in the nucleon.  Using the observed value of
$\sigcom$ \cite{sigcom} gives $S_{CR}=-250$ GeV$^{-2}$, which is comparable
in strength to phenomenological scalar forces. Unlike the total scalar
potential, this strength will not be changed by including $\pi\pi$
interactions, provided the couplings and cut-offs are chosen to reproduce the
observed $\sigcom$. A crude estimate of the symmetry restoring potential in
matter is $\rho S_{CR}$, which is about $-330$ MeV at nuclear matter density,
almost as large as phenomenological scalar potentials
\cite{walecka,relnp,dirph,ray}. The implications of this for a nucleon in
matter will be discussed in Secs.~3.3 and 3.4.

\section{Finite density}
\subsection{Quark condensate}
The vacuum quark condensate is negative (2.15) while the scalar densities of
quarks in hadrons are positive. As a result the net quark condensate will be
smaller in nuclear matter than in vacuum. At low densities we can treat the
nucleons as independent and simply add their contributions to get the spatially
averaged scalar density of quarks:
$$2\mbar\langle \qbar q\rangle_\rho=2\mbar \langle \qbar q\rangle_0
+\sigcom\rhob. \eqno(3.1)$$
By taking the ratio to the vacuum condensate and using the GOR relation (2.15)
we can cancel the poorly known current mass to get the model-independent
result \cite{cfg}
$${\langle \qbar q\rangle_\rho\over \langle \qbar q\rangle_0}=1-{\sigcom\over
f_\pi^2m_\pi^2}\rhob. \eqno(3.2)$$
This was first obtained by Drukarev and Levin in the context of a QCD sum-rule
analysis \cite{dl}, and has also been noted in the NJL and linear sigma models
\cite{lkw,cs1,cs2}. Assuming that this linear density dependence is valid up to
nuclear matter density, $\rhob\simeq 0.17$ fm$^{-3}$, and with $\sigcom=45\pm
7$ MeV \cite{sigcom}, one finds a $\sim 30$\% reduction in the quark condensate
in nuclear matter. Extrapolating to higher densities suggests that chiral
symmetry could be completely restored at about three times the density of
nuclear matter.

A qualitative picture of the effects of finite baryon density can be obtained
 from either the linear sigma model or the NJL model. The latter is more often
used in calculations of symmetry restoration \cite{njlfd,jam,lkw,cfg,cs2}
since it contains a Dirac sea of quarks that can provide a quark condensate.
However we have at present no consistent way to describe nuclear matter within
this model. Many treatments therefore take the rather drastic step of replacing
nuclear matter by a uniform Fermi gas of quarks. This neglects the strong
correlations in real nuclear matter that cluster the quarks in threes to form
colour-singlet nucleons and then tend to keep those clusters apart. Such
correlations lead to partial occupation of many more quark states \cite{brw}.
Hence a degenerate Fermi gas grossly overestimates the Pauli blocking of quarks
in matter. Jaminon \etal \cite{jam} have suggested a hybrid approach, treating
the vacuum as a Dirac sea of quarks but using a Fermi sea of nucleons. Although
not fully consistent, this approach is probably more realistic than ones based
on degenerate quark matter.

In either the linear sigma model or a bosonised NJL model, the energy density
for uniform matter can be written
$${\cal E}(\sigma)=U(\sigma)+{\gamma\over(2\pi)^3}\int_0^{k_F}\!d^3{\bf
k}\,\sqrt{{\bf k}^2+M^2}, \eqno(3.3)$$
where $U(\sigma)$ is a Mexican hat potential, the fermion mass is $M=g\sigma$,
and the degeneracy factor is $\gamma=4$ for a gas of nucleons and $\gamma=12$
for one of quarks. These contributions to the energy density in the linear
sigma model are shown in Fig.~3.1. For large values of $\sigma$ or low
densities the fermion energy is linear in the baryon density $\rhob$:  $${\cal
E}(\sigma)\simeq U(\sigma)+\rhob g|\sigma|. \eqno(3.4)$$ At low densities this
produces a reduction in the vacuum value of $\sigma$, and hence in the quark
condensate, which is also linear in the density.  The consequences of this
include reductions in the nucleon mass and the pion decay constant in matter.

As the density increases further the scalar field is pushed to smaller values
where the fermions start to become relativistic and so couple less strongly to
$\sigma$. Finally a density is reached where the maximum at the centre of the
Mexican hat is replaced by a minimum. In the case of a quark gas in the absence
of explicit symmetry breaking, this corresponds to a second-order phase
transition above which chiral symmetry is restored, the condensate vanishes
and fermions are massless. When current masses are included, there is no actual
phase transition: the condensate and masses go rapidly but smoothly to small
values. For the low sigma masses typical of the NJL model this happens at about
three times the normal density of nuclear matter. In contrast, a gas of
nucleons remains nonrelativistic to higher densities, and so couples more
strongly to the sigma field. The stronger coupling enhances the nonlinear
density dependence, leading to an even lower critical density and a transition
which tends to be first order.

These typical features are illustrated in Fig.~3.2 for quark and nucleon Fermi
gases in the linear sigma model; they can also be seen rather clearly in the
work of Jaminon \etal \cite{jam} for the NJL model. In both quark and nucleon
cases the quark condensate initially decreases linearly with density. At higher
densities this decrease becomes more rapid and, for nucleons, a Lee-Wick phase
transition \cite{leewick} occurs at about half nuclear-matter density. The
behaviour of the quark curve may look more plausible, but one should remember
that it describes degenerate quark matter, not nuclear matter. Although such
models can provide qualitative sketches of chiral symmetry restoration, their
details should not be taken seriously. In particular repulsive forces, such as
$\omega$ exchange, have not been included and use of a light $\sigma$ meson
substantially overestimates the nonlinear density dependence.

At low densities the behaviour of the quark condensate is given by (3.2),
independently of whether a gas of nucleons or quarks is used. The
model-independent nature of this result was pointed out by Cohen, Furnstahl and
Griegel \cite{cfg}, who obtained it using the Feynman-Hellmann theorem. This
derivation is worth pursuing because it provides a connection between the quark
condensate in matter and the interactions between the nucleons. It can
therefore be applied to more realistic models of nuclear matter than those
discussed so far. The energy density of nuclear matter can be written in the
form
$${\cal E}={\cal E}_0+\MN\rhob+\delta{\cal E}, \eqno(3.5)$$
where ${\cal E}_0$ is the vacuum energy density (independent of $\rhob$) and
$\delta{\cal E}$ includes terms of higher order in $\rhob$, coming from nucleon
kinetic energies and potentials. The scalar density of quarks can found by
differentiating (3.5) with respect to $\mbar$ and applying the Feynman-Hellmann
theorem (2.17). This gives (3.1) plus higher-order corrections arising from
the binding energy in (3.5). Since the binding energy per nucleon is less than
2\% of $\MN$ at the density of nuclear matter, these higher-order corrections
are expected to be small.

This expectation is borne out by estimates of the density dependence of the
quark condensate in several models of nuclear matter. Chanfray and Ericson
\cite{ce} have looked at pionic effects on the sigma commutator in matter. They
find a strong cancellation between Pauli blocking of the pion cloud and pion
exchange incorporating tensor correlations. This leaves a small ($<10$\%) net
enhancement of the symmetry restoration over (3.2). Similar enhancements
are found from estimates of contributions from heavier-meson exchanges
\cite{cfg,restore}. In all these models the higher-order effects are rather
smaller than those in the linear sigma and NJL models mentioned above.

Although the quark condensate in matter can always be calculated directly
 from a model wave function, it can also be obtained by taking the soft-pion
limit of the pion propagator \cite{ca,reya}. This is analogous to the soft-pion
theorems leading to the GOR relation (2.14) and connecting $\sigcom$ to $\pi N$
scattering. M. Ericson \cite{ericson} has suggested that the condensate can
include not just meson-exchange effects but also a ``distortion factor," coming
 from rescattering of soft pions in the nuclear medium. Such a factor would
tend to reduce the amount of chiral symmetry restoration. However, PCAC imposes
relations between the various contributions to the scattering amplitude which
ensure that only the symmetry-breaking matrix element survives at the soft
point. All such contributions of a given order in the density must be included
to be consistent with the constraints of PCAC. Also the pion propagator must of
course be defined using the PCAC interpolating field (2.10). The sigma
commutator per nucleon evaluated in this way should agree with a direct
calculation of it as a symmetry-breaking matrix element without reference to
any scattering process. The rescattering term \cite{ericson} does not satisfy
these conditions \cite{restore}.

These features can be illustrated by a calculation of soft-pion scattering to
second order in the density $\rho$ in the linear sigma model \cite{restore}. To
first order in the density, the scattering amplitude per unit volume is just
$\rho$ times the free-nucleon result (2.32). Strictly, the leading correction
to this arises from the Fermi motion of the nucleons and is of order
$\rho^{5/3}$ but at normal nuclear densities this term is negligible. The most
important corrections are thus of second order in the density. For simplicity
these can be calculated in the static approximation, neglecting the motion of
the nucleons. These second-order contributions include the rescattering term
\cite{ericson}, which can be represented diagramatically in Fig.~3.3(a).
However one must also include the one-pion-irreducible (OPI) diagrams for pion
off two nucleons. The importance of keeping all terms at a given order in the
density has long been known in the context of pion-exchange effects on
pion-nucleus scattering \cite{robwil}. Indeed the cancellations between such
terms are essential if the pion-exchange contribution to the sigma commutator
is to be obtained in the soft-pion limit \cite{ce}.

Working at tree level the OPI contributions are given by the diagrams of
Figs.~3.3(b-d). These are the $\sigma$-exchange corrections to the lowest-order
result. Other two-body diagrams, for example pion-exchange terms, do not
contribute for static nucleons. The soft-pion scattering amplitude including
all the diagrams of Fig.~3.3 corresponds to a scalar quark density of
$$2\mbar\langle \qbar q\rangle_\rho-2\mbar \langle \qbar q\rangle_0
=\sigcom\rho+{3\over 2}{\sigcom^2\over f_\pi^2m_\pi^2}\rho^2.
\eqno(3.6)$$
For comparison, the rescattering diagram, Fig.~3.3(a), gives an order-$\rho^2$
correction of similar size but with a coefficient of $-1$ instead of ${3\over
2}$. The net effect at this order is thus an enhancement of symmetry
restoration in matter, rather than a reaction against it as suggested by
rescattering alone. The scalar density (3.6) agrees with a direct evaluation of
the matrix element of the symmetry-breaking term in the model (2.28), as it
should because the pion field of the linear sigma model is the interpolating
field of PCAC.

It is instructive to rederive this result (3.6) using the Feynman-Hellmann
theorem since this makes it clear why $\sigma$ exchange acts to enhance the
sigma commutator. The energy density for this simple model of static nucleons
interacting via $\sigma$-exchange is
$${\cal E}=\MN\rho-\half {g^2\over m_\sigma^2}\rho^2. \eqno(3.7)$$
For fixed $\lambda$ and $\nu$ the relations (2.29, 30) can be used to evaluate
the derivatives of the nucleon and sigma masses with respect to $m_\pi^2$. With
these, the derivative of (3.7) yields the same result for the change in the
quark condensate as (3.6). Sigma exchange is an attractive interaction whose
strength is increased as we switch off the symmetry breaking ($m_\pi\rightarrow
0$). It therefore tends to enhance the sigma commutator in matter.

Taking $\sigcom=45$ MeV \cite{sigcom} in (3.6) would suggest that $\sigma$
exchange increases the sigma commutator per nucleon by about 24 MeV at nuclear
matter density. Of course, one should also include other components of the $NN$
interaction. For example, $\omega$ exchange is a repulsive force whose strength
also increases as $m_\pi\rightarrow 0$ \cite{cfg}, tending to counteract the
effect of $\sigma$ exchange. A simple estimate suggests that it will produce a
reduction in the sigma commutator of about 10--15 MeV at nuclear matter
densities. Short-range correlations between the nucleons will tend to reduce
both of these meson-exchange contributions, leaving rather small deviations
 from the linear density dependence of the quark condensate.

The repulsive short-range correlations between nucleons obviously cut down the
contributions of heavy meson exchanges to the quark condensate. In soft-pion
scattering it might look as though they could reduce the $\sigma$-exchange
effects while leaving the rescattering term of Ref.~\cite{ericson} unchanged.
However T. Ericson has shown that strong correlations must also affect the
rescattering \cite{te}. In a toy model of isolated static nucleons, whose
separations are larger than the range of the $\pi N$ interaction, the
propagation of the pions in the empty space between the nucleons can be
expressed in terms of on-shell pions only\cite{beg}. This remains true even if
one even extrapolates the scattering amplitude so that the incident pion
momentum is off-shell. The initial scattering in any term of the soft-pion
amplitude thus involves one soft and one on-shell pion. If the PCAC field has
been used to define the off-shell extrapolation, then the amplitude is zero by
the Adler condition \cite{ca,reya}. Hence in the limit of extreme correlations,
both the rescattering and the OPI diagrams of Fig.~3.3 vanish. This is of
course consistent with the vanishing of the $\sigma$ exchange contribution to
$\sigcom$ in this limit.

It has been pointed out \cite{te} that strong correlations between nucleons
could have another equally important consequence if changes in the quark
condensate were short ranged. Each nucleon would then be an island of symmetry
restoration surrounded by normal vacuum and the spatial average of the quark
condensate in matter, described by (3.2), would not be a very relevant
quantity: the  repulsion between nucleons would mean that quarks of a given
nucleon would not feel the chiral symmetry restoration produced by its
neighbours. However this picture is not realised since, as described in
Sec.~2.6, a major contribution to this restoration arises from two-pion
exchange between the nucleons \cite{btwo}. This has a moderately long range,
similar to that of the scalar attractive force, even if the ``elementary"
$\sigma$ meson has a large mass. Consequently symmetry restoring effects should
not be dramatically suppressed by the short-range correlations between nucleons
and partial symmetry restoration could have significant effects on nucleons in
nuclei. For instance, the symmetry restoring potential experienced by a nucleon
is not expected to be substantially reduced from the estimate of $-330$ MeV in
Sec.~2.6.

The estimates of chiral symmetry restoration just discussed are all based on
simplified treatments of the $NN$ interaction and nuclear matter. To improve on
these we need calculations based on realistic forces with correlations between
nucleons obtained consistently from these forces, for example using a
relativistic Brueckner-Hartree-Fock approach \cite{relnp,rbhf}. The
Feynman-Hellmann approach could be applied to estimate the corresponding quark
condensate provided one can make reasonable assumptions about the dependence on
$\mbar$ (or $m_\pi^2$) of the masses and couplings of the exchanged mesons. It
should also be possible to estimate how much symmetry restoration a nucleon
experiences in matter, taking into account correlations.

A large change in the quark condensate in matter is important since it may lead
to comparable modifications of hadron properties. In particular there would be
significant effects on meson and baryon masses.  The effects on meson masses
and decay constants are described in Sec.~3.2. The masses of pions and kaons
are discussed in some detail because of suggestions that kaon condensation
could
occur in dense nuclear matter. Effects on baryon properties have been
estimated using a variety of models and these are summarised in Sec.~3.3.
Possible observable consequences of these changes in masses and couplings are
described later, in Sec.~4. Applications of QCD sum rules to hadrons in matter
are illustrated by the calculation of the nucleon scalar and vector
self-energies (Sec.~3.4). Possible changes in hadron properties arising from
changes in the gluon condensate are discussed in Sec.~3.5.

\subsection{Mesons in matter}
In a naive quark-model picture, with constituent quark masses generated
dynamically, one would expect the masses of non-strange mesons (except for the
pions) and baryons to decrease by a similar amount, related to the quark
condensate. Since pions would be Goldstone bosons in the chiral limit their
masses should remain small, at least until the symmetry-restoring phase
transition. When the symmetry is restored each state should become degenerate
with a chiral partner of the opposite parity.

Calculations in the NJL model display this expected behaviour for meson masses,
although one should not take seriously any results obtained using a Fermi gas
of quarks \cite{njlfd,lkw}. The Pauli blocking in such a model strongly affects
$s$-wave $q\qbar$ states like the vector mesons. Instead of decreasing as the
density increases, their masses behave like that of the pion, remaining rather
constant at low density and then rising to meet the masses of their
axial-vector chiral partners. The rise of the pion mass at high densities is
similarly an artefact of the model. Such behaviour is not seen when a gas of
nucleons is used \cite{jam}: there the masses of the vector, axial-vector and
scalar mesons all decrease with density until chiral symmetry is restored.

QCD sum rules have also been used to study the masses of the $\rho$ meson in
matter \cite{hl,ak}. A basic problem for an predictions from this approach is
that the four-quark condensate $\langle(\qbar q)^2\rangle$ plays a dominant
role in the sum rule. As mentioned in Sec.~2.5, this condensate is not well
determined. Even if the factorised ansatz (2.50) is valid in the vacuum, it is
not at all clear that this continues to hold in matter. For instance, at finite
temperature the contributions from the pion gas do not have this form
\cite{hkl}. The results for the medium dependence of $m_\rho$ are thus
crucially dependent on the assumed behaviour of the four quark condensate. If
this condensate is taken to decrease strongly with density, as suggested by the
factorised ansatz and (3.2), then a strong decrease of $m_\rho$ is found
\cite{hl,ak}. This is sufficient to overcome the tendency for the mass to
increase because of the effect of $\Delta$-hole excitations on the $\pi\pi$
component of the $\rho$ \cite{aklq,hfn}.

As long as the chiral symmetry remains hidden, the pion is an approximate
Goldstone boson and large changes in its mass are not expected. The linear
dependence on the density of the pion mass can be found from the isoscalar
$\pi N$ scattering amplitude at threshold, usually expressed in terms
of a scattering length, $a^{(+)}=-0.010m_\pi^{-1}$ \cite{ew}. This scattering
length is of order $\mbar$, like $m_\pi^2$ in free space, but is very small
even by comparison with other quantities of this order. For example it
corresponds to a scattering amplitude at threshold that is about five times
smaller than at the soft point (2.32). Delorme, Ericson and Ericson \cite{dee}
(see also \cite{ynk}) have pointed out that this gives a very weak density
dependence of the pion mass in matter. Similarly the NJL model predicts
remarkably small changes in the pion mass at nuclear densities \cite{lkw,jam}.
Lutz \etal \cite{lkw} have interpreted this in terms of a screening of the
scalar interactions, resulting from the finite size of the quark
``quasi-particles" in this model. The pion mass thus has an additional
protection against modification, beyond that expected from chiral symmetry. In
particular there is no evidence for its rapid decrease, leading to $s$-wave
pion condensation, as has been suggested \cite{bkr}.

The behaviour of kaons in matter is less well understood. There have been
repeated suggestions that $s$-wave kaon condensation could occur at a few times
normal nuclear densities, which could have important consequences for
supernovae and the formation of neutron stars \cite{bb,tpl}.

Early suggestions \cite{kapnel,nelkap,clean,pw} relied on a scalar attraction
between the kaon and nucleon, driven by a large kaon-nucleon sigma commutator.
That requires a large strangeness content of the nucleon which, as discussed in
Sec.~2.2, is not realistic. Any such attraction is further reduced by the
momentum dependence of the KN scattering amplitude, as the pion case
\cite{dee,ynk}. However the idea is not dead because other sources of
attraction have been found. In particular the Weinberg-Tomozawa \cite{ca} term
provides a current-current interaction that is attractive in $K^-N$ scattering
\cite{nelkap,pw,bkrt,lsw}. This interaction is often modelled by $\rho$ and
$\omega$ exchange in phenomenological treatments \cite{knint}. In the $K^+N$
case it is repulsive and so there is no tendency for $K^+$ condensation to
occur \cite{lsw}.

Further effects that can promote the formation of a $K^-$ condensate have
been pointed out by Brown \etal \cite{bkrt,blrt}. If a Fermi gas of electrons
is present, as in neutron star matter, then a condensate can form when the
$K^-$ rest energy drops to the electron chemical potential. The symmetry
energy of nuclear matter favours the conversion of neutron matter to nuclear
matter with a $K^-$ condensate and so lowers the critical density for
condensation.

One approach to this problem \cite{pw,bkrt,blrt} makes use of effective chiral
Lagrangians from ChPT \cite{jenman,chptrev}. The most recent calculations start
 from a Lagrangian that includes terms up to third order in $m_\pi$ or the
small momenta of the chiral expansion \cite{ljmr,lbr}. A $\Lambda$(1405) field
is added to the model to account for the rapid energy dependence of $K^-p$
scattering in the vicinity of that resonance. With an attractive force between
the $\Lambda$(1405) and the nucleon, the model can provide sufficient
attraction for $K^-$ condensation at about four times nuclear matter density
\cite{lbr}. However the convergence of the chiral expansion seems rather slow
\cite{ljmr}, as might be expected from the large kaon mass.

Other approaches are based on scattering amplitudes defined in terms of the
PCAC interpolating kaon field \cite{lsw,ynk,ynmk}. These do not find sufficient
attraction for kaon condensation. The difference from the ChPT results is not
due to the choice of interpolating field: that is purely a matter of
convenience and no observable quantity can depend on it. Rather it arises
because the two models contain different dynamics at second order in the
density \cite{ymk}. For the approaches to be consistent, they need to
include six-point interactions describing kaon scattering from two nucleons.
Such terms give rise to irreducible contributions of order $\rho^2$ to the kaon
self-energy in matter, analogous to those discussed in Sec.~3.1 for the pion
case \cite{restore}. Moreover there are many such terms of second order in the
chiral expansion of the $K$-nucleus scattering amplitude; current ChPT
calculations are thus incomplete at that order.

To decide whether there is sufficient attraction for kaon condensation, it is
clearly essential to have empirical input to tie down the density dependence of
the $K^-$ interaction with nuclear matter. The scattering lengths, which
control the low density behaviour, are of little help: the $K^-n$ scattering
length is poorly determined \cite{barswa}, while the $K^-p$ one is complicated
by the $\Lambda$(1405) resonance just below threshold. The $K^+N$ scattering
lengths can at least provide some check on the models used. Perhaps the best
sources of such information are kaonic atoms. A recent analysis \cite{fgb} of
these using a density-dependent $K^-$ optical potential does find a substantial
$K^-N$ attraction \cite{fgb}, although this cannot be reliably extrapolated
to the densities of interest for kaon condensation.

As well as modifying mesons' masses, nuclear matter is expected to alter their
decay constants. In the present context the pion decay constant is of
particular interest since it is directly related to the hidden nature of chiral
symmetry in the QCD vacuum, as indicated by its definition (2.8). Any decrease
in $f_\pi$ in matter therefore provides a signal of partial symmetry
restoration. Indeed such a reduction is found in both linear sigma
\cite{leewick,akhm,abpw,cek} and NJL models \cite{njlfd,lkw}. By altering the
induced pseudoscalar coupling constant in nuclear matrix elements of the axial
current, changes in $f_\pi$ can have observable effects on rates of muon
capture \cite{mucap}.

\subsection{Nucleons in matter}
Partial restoration of chiral symmetry can also produce changes in nucleon
properties. Unfortunately we have as yet no consistent model that describes
both the quark structure of nucleons and the binding of those nucleons to form
nuclei. A hybrid approach is often used where a quark- or soliton-model nucleon
is embedded in mean fields taken from a model for nuclear matter. Alternatively
QCD sum rules can be applied to a nucleon propagator in matter. Possible
observable consequences of the changes in nucleon properties suggested by these
models will be discussed in Sec.~4.

Some authors have taken a density-dependent pion decay constant from the NJL
model and used this to construct solitons of either a Skyrme \cite{mmsk} or
linear sigma model \cite{acg}. Christov and Goeke \cite{christov} have studied
the properties of an NJL soliton embedded in a Fermi gas of quarks. Another
approach is based on either a bag \cite{guichon} or nontopological soliton
model \cite{mmsol} where the quarks are coupled to scalar and vector
fields. Despite their differences, all of these models yield qualitatively
similar results.

The nucleon mass is found to decrease in matter by about 15--20\%. Although
significant, this is much smaller than the $\sim40\%$ reduction found in, for
example, the $\sigma$-$\omega$ model \cite{walecka}. Charge radii increase by
10--20\% together with magnetic moments, and so partial symmetry restoration
can provide a mechanism for the often-suggested ``swelling" of nucleons in
matter \cite{swell}. The coupling of the nucleon to the axial current $g_A$
decreases slightly ($\sim 5$\%), as a result of the quarks becoming more
relativistic as their mass decreases. That also reduces the coupling of scalar
fields to the nucleon \cite{guichon,mmsol}.

The QCD sum rule approach described in the following section can also lead to a
decrease of the nucleon mass with density, although this conclusion is
sensitive to assumptions about the behaviour of the four-quark condensate. Both
this approach and the models mentioned above have been used to study the
difference between proton and neutron masses in matter; this will be discussed
further in Sec.~4.3. Other nucleon properties at finite density have not been
studied at finite density using QCD sum rules, although Henley \etal \cite{hhk}
have pointed out that a sum rule for $g_A$ does suggest that this quantity will
decrease if chiral symmetry is restored.

\subsection{QCD sum rules}
QCD sum rules are an alternative to models, which  may provide a more direct
way to relate changes in nucleon properties to changes in the various
condensates in matter.  Drukarev and Levin \cite{dl} have applied the sum rule
approach to the problem of nuclear binding and saturation. However, as pointed
out by Cohen, Furnstahl, Griegel \etal \cite{cfgsr} (hereafter denoted by
CFG$+$), the sum rule method is not precise enough to make meaningful
predictions for such quantities. Instead that group have looked at the
self-energy of a nucleon in matter. Henley and Pasupathy \cite{hp} have carried
out a similar calculation, expanding both sides of the sum rules to first order
in the density.

In matter the Green's function (2.41) for the nucleon interpolating field
can be written in terms of three invariant functions:
$$\Pi(q)=\Pi_s(q^2,q\cdot u)+{q\llap/}\Pi_q(q^2,q\cdot u)
+{u\llap/}\Pi_u(q^2,q\cdot u), \eqno(3.8)$$
where $u$ is the four-velocity of the matter. It is convenient to work in the
rest frame of the matter where $q\cdot u=q_0$ is just the energy. CFG$+$ point
out that nucleon and antinucleon propagation in matter are not simply related
by charge conjugation. The nucleon pole becomes somewhat broadened in matter as
a result of coupling to two-particle-one-hole states, for example. Nonetheless
it is still narrow on hadronic scales and so can be reasonably well
approximated by a single pole. On the other hand the antinucleon can annihilate
in matter and becomes a very broad structure. CFG$+$ therefore suggest working
with a spectral representation for the propagator in terms of $q_0$, and
choosing the weighting function to emphasise the quasi-nucleon pole at positive
energy.

The covariant form for the propagator of a nucleon in the presence of scalar
and vector potentials leads to the following expressions for the invariant
functions in (3.8):
$$\Pi_s(q_0,|{\mbold q}|)=-\lambda_N^{*2}{\MN^*\over (q_0-E_q)(q_0-\overline
E_q)}+\cdots, \eqno(3.9{\rm a})$$
$$\Pi_q(q_0,|{\mbold q}|)=-\lambda_N^{*2}{1\over (q_0-E_q)(q_0-\overline
E_q)}+\cdots, \eqno(3.9{\rm b})$$
$$\Pi_u(q_0,|{\mbold q}|)=+\lambda_N^{*2}{\Sigma_V\over (q_0-E_q)(q_0-\overline
E_q)}+\cdots, \eqno(3.9{\rm c})$$
where the dots denote continuum contributions and polynomial terms. This has
positive- and negative-energy poles at
$$E_q=\Sigma_V+\sqrt{{\mbold q}^2+\MN^{*2}}, \eqno(3.10{\rm a})$$
$$\overline E_q=\Sigma_V-\sqrt{{\mbold q}^2+\MN^{*2}}, \eqno(3.10{\rm b})$$
where $\MN^*$ is a nucleon's (Dirac) mass in matter and $\Sigma_V$ is its
vector potential. The residue $\lambda_N^{*2}$ describes the coupling of the
interpolating field (2.44) to the quasi-nucleon.

To suppress the antinucleon contribution, CFG$+$ split each invariant function
into pieces that are even and odd in $q_0$:
$$\Pi_i(q_0,|{\mbold q}|)=\Pi_i^E(q_0,|{\mbold q}|)+q_0\Pi_i^O(q_0,
|{\mbold q}|). \eqno(3.11)$$
They then apply a Borel transform in $q_0$ at to the combination
$\Pi_i^E(q_0,|{\mbold q}|)-\overline E_q\Pi_i^O(q_0,|{\mbold q}|)$ at fixed
$|{\mbold q}|$. This is equivalent to a weighting function with a factor
of $q_0-\overline E_q$ so that the contribution from the negative-energy
pole is removed. The Borel transform is similar to that in (2.42, 43) except
that $Q^2=-q_0^2$. This choice ensures that the sum rules reduce to their
vacuum forms at zero density.

For large values of the Borel mass $M$ the transform emphasises large negative
$q_0^2$ with fixed $|{\mbold q}|$. This corresponds to a region of large,
space-like four-momenta where the OPE can be applied. That expansion can be
done in a similar manner to the vacuum case; details can be found in the work
of CFG$+$ \cite{cfgsr}. The main difference from the vacuum OPE is the
appearance of new condensates, such as the dimension-three vector quark
condensate,
$$\langle q^\dagger q\rangle_\rho={3\over 2}\rhob. \eqno(3.12)$$
Other condensates, such as $\langle\qbar q\rangle$, are replaced by their
values in matter.

A simplified version of the sum rules, analogous to (2.48), which illustrates
their main features is
$$\lambda_N^{*2}\MN^*\exp[-(E_q^2-{\bf q}^2)/M^2]=-{M^4\over
4\pi^2}\langle\qbar q\rangle_\rho+\cdots, \eqno(3.13{\rm a})$$
$$\lambda_N^{*2}\exp[-(E_q^2-{\bf q}^2)/M^2]
={M^6\over 32\pi^4}+\cdots, \eqno(3.13{\rm b})$$
$$\lambda_N^{*2}\Sigma_V\exp[-(E_q^2-{\bf q}^2)/M^2]={2M^4\over
3\pi^2}\langle q^\dagger q\rangle_\rho+\cdots. \eqno(3.13{\rm c})$$
The dots denote condensates of dimension four or higher and continuum
contributions; the full forms are given by CFG$+$. Taking ratios of these sum
rules one gets a modified version of the Ioffe sum rule (2.49) for the nucleon
mass,
$$\MN^*=-{8\pi^2\over M^2}\langle \qbar q\rangle_\rho, \eqno(3.14)$$
and a vector self-energy given by
$$\Sigma_V={32\pi^2\over M^2}\rhob. \eqno(3.15)$$
The qualitative features of these sum rules are consistent with relativistic
phenomenology \cite{walecka,relnp,dirph,ray}. The change in the scalar
condensate (3.2) drives a reduction in the nucleon mass and so provides a
scalar attraction, while the vector condensate produces a repulsive vector
self-energy.

CFG$+$ have studied the effects of higher condensates omitted from (3.13) and,
with one exception, they find their results to be insensitive to them. The
exception is the four-quark condensate $\langle(\qbar q)^2\rangle_\rho$ which,
as we saw above, plagues the $\rho$-meson sum rule too. If that condensate is
assumed to vary weakly with density the expectations based on the simplified
sum rules are fulfilled: in nuclear matter the nucleon mass is reduced to
roughly 60\% of its free-space value and there is a vector self-energy of about
300 MeV. On the other hand, if the condensate is given a strong density
dependence, as suggested by the factorised ansatz, then the nucleon mass
remains almost unchanged in matter. Since the vector self-energy is still
large, this gives a quasi-nucleon energy that is substantially larger than
$\MN$. Such a situation seems unrealistic. However it is clear that more work
is required to determine the density dependence of the four-quark condensate
$\langle(\qbar  q)^2\rangle_\rho$ before QCD sum rules can yield reliable
predictions for mesons and baryons in matter.

\subsection{Scaling?}
Partial restoration of chiral symmetry is expected to decrease hadron masses in
matter. Linear sigma \cite{leewick,akhm,abpw} and NJL models \cite{njlfd} also
show a reduction in the pion decay constant. The results of such models have
led to the suggestion \cite{brown} that quantities such as the nucleon,
$\sigma$ and vector meson masses and $f_\pi$ all behave similarly:
$${\MN^*\over \MN}\simeq{m_\sigma^*\over m_\sigma}\simeq{m_V^*\over m_V}\simeq
{f_\pi^*\over f_\pi}, \eqno(3.16)$$
where the stars denote values in matter. Brown and Rho \cite{brscale} (see also
\cite{barev}) have extended this idea by proposing that there is a single
relevant length scale in nuclear matter, essentially $f_\pi^*$, and have
suggested that this might be a consequence of the broken scale invariance of
QCD, which leads to a single dimensioned parameter $\Lambda_{\rm QCD}$ in the
theory (apart from current quark masses).

The QCD scale anomaly \cite{scanom} can be incorporated in low-energy effective
Lagrangians by adding an extra scalar, isoscalar field, the dilaton
\cite{schdil,gjjs}, whose vacuum expectation value provides the only scale. The
self-interaction potential for this field, denoted by $\chi$, is taken to be of
the form
$$V(\chi)=a\chi^4+b\chi^4\ln(\chi/\chi_0).  \eqno(3.17)$$
The first term provides a scale-invariant classical potential that on its own
would give a vanishing vacuum expectation value for $\chi$. It would also
leave the dilaton excitations massless, rather like Goldstone bosons. The
second term models the quantum effects responsible for the scale anomaly. It
explicitly breaks scale invariance, driving the vacuum to a nonzero value of
$\chi$ and providing a mass for the dilaton excitations. The single dimensioned
parameter of the model is $\chi_0$, which sets the scale of all other
dimensioned masses and couplings. From a scaling of all dimensioned quantities,
one finds that the $\chi$ field can be related to the trace of the
stress-energy tensor by
$$-4b\chi^4=T_\mu^\mu. \eqno(3.18)$$
This trace contains all effects that break scale invariance and in QCD it takes
the form \cite{scanom}
$$T_\mu^\mu=-{9\alpha_s\over 8\pi}G_{\mu\nu}^a G^{a\mu\nu}+m_u\ubar u
+m_d\dbar d+m_s \sbar s.   \eqno(3.19)$$
In the vacuum this is dominated by the contribution of the gluon condensate
$\langle (\alpha_s/ \pi)G_{\mu\nu}^a G^{a\mu\nu}\rangle\simeq (360\pm 20$
MeV)$^4$ \cite{svz,rry,qssr}.

The problem with such an approach is that the scale anomaly of QCD is large and
so the theory is not approximately scale invariant \cite{banrev,birsc}. This
can be seen from the fact that the lightest scalar glueball, which one might
hope to identify with a dilaton, is estimated to lie at around 1.5 GeV
\cite{glueball,morgan}. If the dilaton were light enough compared to other
states in the same channel, the relation (3.18) could be used to define an
interpolating dilaton field by analogy with the pion field of PCAC (2.10). This
could then be used to obtain ``soft dilaton theorems" describing the
consequences of approximate scale invariance which could be embodied in
Lagrangians with a dilaton field. In reality there are many other scalar,
isoscalar states in the energy range 1--2 GeV and so a single pole is most
unlikely to dominate matrix elements of the stress energy tensor. Hence an
interpolating dilaton field, introduced as if it were almost a Goldstone boson,
is not a useful ingredient in low-energy effective Lagrangians for QCD.

Even if a dilaton field is introduced, it remains almost unchanged in hadronic
matter at normal densities. This has been found in many applications of such
models \cite{jjs,ripjam} (see \cite{birsc} for a list of further examples):
significant changes in the gluon condensate are not produced inside hadrons or
normal nuclear matter if realistic values of the glueball mass and gluon
condensate are used.

This stiffness of the gluon condensate is another consequence of the lack of
scale invariance of QCD. It is clearly shown in the work of Cohen, Furnstahl
and Griegel \cite{cfg}, which uses the trace anomaly to relate the change in
the gluon condensate to the energy density of hadronic matter. In stable
nuclear matter the change in $T_\mu^\mu$ is simply the energy density of the
matter since the pressure vanishes. For normal nuclear matter this gives a
change in the gluon condensate of about 150 MeV fm$^{-3}$. This should be
compared with the vacuum gluon condensate of 2200 MeV fm$^{-3}$. Even allowing
for a factor of two uncertainty in this condensate, its change in nuclear
matter is at most a 15\% effect. The fourth root of the condensate, which
corresponds to the change in the dilaton field or the change of scale, is
altered by no more than 4\%.

There are only two ways to get large changes in the gluon condensate at normal
densities relative to its vacuum value. One is to take a value for the
vacuum condensate very much smaller than that the deduced from QCD sum rules.
That would mean rejecting the rather well tested applications of those sum
rules to charmonium \cite{svz,rry,qssr}. The other is to use a $\chi$ field
with a much lighter mass than any scalar meson. The gluon condensate would then
be very soft and so its response could be large. Universal scaling would arise
in such a model, as noted by Kusaka and Weise \cite{kuswei},\footnote{The
scaling hypothesis leads to hadron masses that vary as the cube root of the
quark condensate. Such a relationship has also been found in a version of the
NJL model without taking a very large mass for the scalar meson \cite{ripjam}.
However in that model the relationship between the masses and the quark
condensate is not a consequence of scaling but instead arises from the
artificial choice of a model involving four-body rather than two-body forces
between the quarks.} since the changes in quark condensate generated indirectly
via the gluon condensate would be much larger than those produced directly by
the scalar density of quarks in matter. However this would require a light
dilaton, even though no such particle is observed. It would also be
inconsistent with observed nuclear binding energies, since the scale anomaly
provides a connection between these and the change in the gluon condensate.
Neither of these choices seems acceptable.

The quark and gluon condensates thus behave very differently at finite density
and there are at least two scales relevant to nuclear matter, as recognised
in \cite{barev}. Moreover the stiffness of the gluon condensate means that any
universal scaling is very small at normal nuclear densities. The size of the
$\pi$N sigma commutator and its associated form factor \cite{sigcom} show that
the quark condensate is significantly deformed in the presence of valence
quarks. This can occur even if the ``elementary" scalar meson is heavy because
of its strong mixing with the two pion channel \cite{birse}. Any changes in
hadron properties in matter are thus likely to be consequences not of scaling
but of the partial restoration of chiral symmetry described in Sec.~3.1.

\section{Signals of symmetry restoration}
Various experimental observations have now been cited as signals for the
modification of hadron properties in nuclear matter, consistent with partial
restoration of chiral symmetry. Unfortunately almost none of these provides
unambiguous evidence since there are ``conventional" mechanisms such as
short-range correlations, $\Delta$-hole excitations or meson exchange currents
which can generate similar density-dependent effects. The important exception
is the axial charge, discussed below. A further complication is the fact that
effects arising from nucleon structure can be mimicked by Z-graphs in
treatments based on point-like Dirac nucleons, as mentioned in Sec.~2.3.

\subsection{Axial Charge}
The axial charge operator has long been known as a good probe of chiral
symmetry in nuclei \cite{townrev}. Originally interest focussed on the
one-pion-exchange contribution, whose form is governed by a soft-pion theorem
and which can produce significant enhancement over the one-nucleon piece
\cite{kdr,delorme}. More recently, studies of first forbidden $\beta$-decays of
nuclei in the lead region have indicated enhancements of $\sim 100$\% in the
effective axial charge of a nucleon in matter\cite{warb}. Similar enhancements
of $\sim 80$\% have also been found in the tin region \cite{wt}. Calculations
of the soft-pion term give only about 50\% enhancement \cite{delorme,kirch}.
Contributions from higher-order terms in ChPT have been found to be small and
only weakly dependent on atomic number \cite{park}.

Interest has therefore switched to exchanges of heavier mesons, and in
particular scalar mesons. Delorme and Towner \cite{dt} have pointed out that
Z-graphs involving these can produces a further enhancement in relativistic
treatments of nuclei. Detailed calculations based on realistic NN interactions
have shown that, combined with pion exchange, these effects can explain the
axial charges deduced from first forbidden $\beta$-decays
\cite{krt,towner,wt,wbt}.

Of the heavy-meson contributions, the direct scalar-exchange term is the most
important. Other scalar- and vector-exchange Fock terms, although individually
large, tend to cancel \cite{krt}.  Although Z-graphs are responsible for this
enhancement when nucleons are treated as point-like Dirac particles, the result
is more general than that type of model.  The direct scalar exchange
corresponds to a reduction in the nucleon mass and this enhances the effective
axial charge to
$$g_A^{c*}={M_N\over M_N^*}g_A. \eqno(3.25)$$
Such behaviour has been noted by Kubodera and Rho \cite{kubrho} in the context
of scaling. A similar result is found in a soliton model for a nucleon embedded
in mean scalar and vector fields \cite{birax}. In that model the change in the
nucleon mass is smaller than in approaches based on Dirac nucleons, as
mentioned in Sec.~3.3, and the enhancement of the form (3.25) is similarly
reduced as a result. However this is compensated by an additional enhancement
arising from changes in the nucleon's structure.

The reaction $pp\rightarrow pp\pi^0$ close to threshold probes similar physics
since the one-nucleon piece involves the axial charge of the nucleon. Moreover
isospin considerations mean that the soft pion term does not contribute and the
kinematics makes this reaction particularly sensitive to heavy meson exchanges.
Recent measurements of the total cross section \cite{meyer} find it to be about
five times that expected from the one-nucleon process. Inclusion of a direct
scalar exchange term, analogous to the corresponding effect in the axial
charge, gives good agreement with the data \cite{leerisk,hmg}.

The enhancement of the axial charge indicates that there are large scalar
fields in nuclei and that the nucleon mass is significantly reduced in matter.
Given that partial restoration of chiral symmetry can contribute significantly
to such phenomenological scalar fields, this is strong evidence for such
symmetry restoration.

\subsection {Other signals}
Other suggested evidence is less conclusive because there are conventional
mechanisms which can produce similar effects. For example, the coupling to the
axial current is believed to be quenched in nuclei \cite{oster}, although the
amount of this cannot be determined model-independently \cite{birmil}. Such
quenching occurs in many models for a nucleon in matter discussed in Sec.~3.3
\cite{mmsk,acg,christov,guichon,mmsol}, where the quarks become more
relativistic as their mass is reduced. However it has long been known that the
effective axial coupling is reduced by core polarisation and $\Delta$-hole
effects \cite{townr2,arimarev}. Calculations of those are not sufficiently
accurate to allow one decide whether any intrinsic quenching is also present.
The situation with magnetic moments and $g$-factors is similar
\cite{townr2,bab,trb}.

The quenching of the longitudinal response seen in quasi-elastic electron
scattering \cite{eepr,muldrev} has often been suggested as a signal for a
``swelling" of nucleons in matter \cite{swell}. Similar effects are also seen
in $(e,e^\prime p)$ reactions \cite{eeprp}. The data can be rather nicely
explained by a $\sim 15\%$ increase in the proton's charge radius
\cite{mulders}. In addition, a $\sim25\%$ increase in its magnetic moment can
fit the increased ratio of transverse to longitudinal responses. These
changes are comparable to those found in the models mentioned in Sec.~3.3.

A similar increase in the charge radius can also be produced if vector meson
masses decrease in matter \cite{brlong,sbr}, assuming that a photon couples to
a nucleon at least part of the time via a virtual vector meson. Nucleon
magnetic moments are also enhanced by this mechanism. A further increase in the
magnetic moments arises from the reduction of the effective nucleon mass,
if the moments are inversely proportional to that mass. This gives a
good description of the ratio of of transverse to longitudinal responses
\cite{sbr}. There is no way to clearly distinguish between an intrinsic
``swelling" of nucleons and a decrease in the masses of vector mesons: both
pictures lead to similar observable consequences.

Of course we should remember that a significant fraction of the missing
longitudinal strength must be due to short-range $NN$ correlations, and that
final-state interactions need to be taken into account. Moreover the agreement
of the data with $y$-scaling indicates that one cannot simply rescale the $Q^2$
dependence of the electric form factor by more than a 10\% change in the charge
radius, and that no significant change in the radius for the magnetic form
factor is allowed \cite{muldrev}. However the soliton model calculations of the
Bochum group \cite{bgkn} suggest that ``swelling" is more complicated than just
a simple rescaling of the nucleon form factors; that model gives a good
description of the longitudinal response although not of the transverse one. As
those authors and others \cite{sbr} note, the transverse response is
complicated by the need to include exchange current effects.

Elastic scattering of $K^+$ mesons provides a good probe of the interior of
nuclei since these particles are only weakly absorbed. Measurements of total
cross sections for $K^+$ scattering from $^{12}$C and deuterium \cite{kplus}
find a ratio that is significantly larger than expected from the impulse
approximation. A much better agreement with the data is obtained if the nucleon
radius is increased by 10\% \cite{siegel}. As with $(e,e')$ scattering, a
reduction of vector meson masses can also explain the data \cite{bdsw}. Similar
discrepancies between the impulse approximation and data on intermediate-energy
proton scattering have also been removed by allowing nucleon and vector meson
masses to decrease in matter \cite{bsh}. However Caillon and Labarsouque
\cite{cl} have noted that decreasing the nucleon as well as meson masses could
lead to too large an effect on $K^+$ scattering. There may also be significant
contributions to kaon scattering from pions being exchanged between the
nucleons \cite{kpiex}.

The EMC effect \cite{emc}, seen in deep-inelastic lepton scattering from
nuclei, shows that there are differences between the momentum distribution of
quarks in a nucleus compared with that for a free nucleon. In particular a
depletion of valence quarks is seen for momentum fractions in the region $x\sim
0.3$--0.5. It is now clear that conventional nuclear binding mechanisms cannot
explain the whole effect \cite{emcth}. An increase in the size of the nucleon
makes the momentum distribution for the valence quarks  more sharply peaked and
so reduces the number of high-momentum quarks. Calculations in bag and soliton
models \cite{emcbag} give changes in the quark distributions of nucleons in
medium which can describe the data quite well.

\subsection{Nolen-Schiffer anomaly}
The differences between the energies of mirror nuclei \cite{nsa} have presented
a long-standing problem in nuclear physics, often referred to as the
Nolen-Schiffer anomaly (NSA). (For a review see Ref.~\cite{shlomo}.) An
intriguing possibility is that changes in the quark structure of nucleons could
lead to a reduction of the neutron-proton mass difference in nuclei and so
resolve the anomaly. Henley and Krein \cite{henkre} estimated the effect of
partial symmetry restoration on the mass difference using an NJL model for the
up- and down-quark masses combined with a nonrelativistic quark model. They
found a decrease of the mass difference with density which was sufficient to
explain the NSA. Similar results have been obtained by other authors taking
different  versions of the NJL model \cite{adbr,llw} or a bag model embedded in
mean scalar and vector fields \cite{saitho}. Hatsuda \etal \cite{hhp}, and
others \cite{adbr,skb}, have applied QCD sum rules to the problem and also
found a decrease in the mass difference. However other models give no effect,
or even an increase in the neutron-proton mass difference at finite density
\cite{meiwei,fcnbg}.

There may also be significant contributions to the NSA from
charge-symmetry-breaking forces, predominantly arising from $\rho$-$\omega$
mixing \cite{bliq}. Sch\"afer \etal \cite{skb} suggest that such effects may be
incorporated in the QCD sum rule approach by including isospin breaking in the
vector self-energy of a nucleon in matter. However such explanations are called
into question by indications that $\rho$-$\omega$ mixing may vary rapidly as
the mesons are taken off-shell \cite{offrho} and hence this mechanism may
contribute much less than expected from the on-shell mixing strength.

There are further complications in the determination of the NSA from the
energies of mirror nuclei. Cohen \etal \cite{cfb} have shown that any
explanation for the NSA based on the local nuclear density leads to a
characteristic pattern for nuclei on either side of a closed shell. Such a
pattern of shell effects is seen in some extractions of the NSA \cite{sato},
but not in others \cite{shlomo}. Other caveats concerning attempts to explain
the NSA have been pointed out by Auerbach \cite{auerb}. At present it is thus
premature to conclude that the NSA is a signal of medium modifications of
nucleons; more consistent treatments of both nucleon and nuclear structure are
needed.

\subsection{Nuclear forces}

Changes in either nucleon structure or meson masses will of course affect the
forces between nucleons in nuclei. This is unlikely to alter our
nonrelativistic pictures of nuclear structure since phenomenological three-body
forces are already included \cite{three}. In fact Hosaka and Toki
\cite{gmatrix}
have shown that density-dependent masses gives G-matrix elements which agree
fairly well with empirical ones.

On the other hand, relativistic treatments of nuclei involve strong scalar
and vector fields which tend to cancel \cite{walecka}. Even rather modest
changes in meson masses or couplings could produce large changes. For example,
a $\sigma$ mass which decreases with density can prevent nuclear matter from
saturating. We may therefore have to look for new mechanisms for saturation,
such as a decrease in the $\sigma N$ coupling due to changes in the quark
structure of the nucleon \cite{guichon,mmsol}.

A large reduction in the nucleon mass leads to an enhanced spin-orbit force,
one of the successes of the Dirac treatment of nuclei \cite{walecka}. The
changes in the nucleon mass indicated by the models mentioned above are less
dramatic. However either a swelling of nucleons or a reduction of vector meson
masses can produce the required additional increase in the spin-orbit force.

A decrease in the $\rho$-meson mass would also affect the tensor force between
nucleons. It would increase the strength of $\rho$-exchange tensor interaction,
leading to more cancellation with pion-exchange and hence a reduction of the
net isovector tensor force \cite{tensor}. Measurements of polarisation
transfer observable in quasi-elastic proton scattering and $(p,n)$ reactions
\cite{poltr} provide some evidence for such a reduction of the tensor and
enhancement of the spin-orbit interactions in nuclei \cite{hls}.

\section{Summary}
Chiral symmetry is partially restored in nuclei: the model-independent result
(1.1) indicates a $\sim 30$\% reduction in the average quark condensate in
nuclear matter. Corrections of higher order in the density are expected to be
small, since they are related to nuclear binding energies. They have been
estimated for both pion and heavier meson exchanges. Although more complete
calculations based on realistic $NN$ forces and including correlations are
needed, these estimates indicate that such corrections are indeed small.

The importance of the pion cloud for the scalar quark density of a single
nucleon implies that two-pion exchange forms a major part of the symmetry
restoration experienced by a nucleon in matter. This suggests a close relation
between symmetry restoration and phenomenological scalar fields in nuclei,
which are often used to model the attractive force produced by two-pion
exchange. The range of two-pion exchange means that, despite the strong
correlations between them, nucleons will experience significant symmetry
restoration in nuclear matter. Again, calculations including realistic $NN$
correlations are called for.

Consequences of the partial symmetry restoration include decreases in both
meson and nucleon masses. Other effects include a reduction in the pion decay
constant. Changes to nucleon properties have been estimated in various models,
although one should bear in mind that none of these provides a consistent
description of nuclei at the quark level. Qualitatively at least the models
are in agreement. As the quark condensate decreases nucleon radii and magnetic
moments increase, while their axial-current coupling tends to decrease. These
changes are driven directly by the reduction in the quark condensate; any
universal scaling of hadron properties, related to the gluon condensate via the
scale anomaly of QCD, is negligible at normal nuclear densities.

Although pions and kaons, as approximate Goldstone bosons, are expected to
behave rather differently from other hadrons, it has been suggested that
attractive forces related to the scalar density could lead to $s$-wave
condensation of these mesons at densities a few times that of normal nuclear
matter. For the pions and $K^+$ this has been ruled out, but for the $K^-$ it
remains a possibility. Kaonic atoms do provide evidence for a strong
$K^-$-nucleus attraction. However present models which attempt to extrapolate
above nuclear-matter density are incomplete and so no definite conclusions can
be drawn.

The effects of symmetry restoration on nucleon properties are consistent
with a number of experimental observations, although almost none of these
provides an unambiguous signal. In general other, more conventional, mechanisms
also contribute and it is hard to disentangle any intrinsic change in nucleon
couplings from them. Examples of such signals, which have been widely touted as
arising from medium modifications but whose interpretation is still unclear,
include the quenched longitudinal response seen in quasi-elastic electron
scattering, the enhanced $K^+$-nucleus total cross sections and the
Nolen-Schiffer anomaly. The one exception is the axial charge, whose
enhancement
shows that there are strong scalar fields in nuclei.

Modifications of hadron properties in matter could significantly alter the
forces between nucleons, in particular strengthening the spin-orbit and
weakening the tensor $NN$ interactions in nuclei. Effects consistent with such
changes have been seen in polarisation transfer observables in proton
scattering and $(p,n)$ reactions.

Further work is needed to improve our theoretical models for the structure
of nucleons in nuclei, and to clarify the role of short-range correlations
and $\Delta$-hole excitations in the electromagnetic response of nuclei.
On the experimental side more information from quasi-elastic electron
scattering is needed to help us understand the missing longitudinal strength.
Changes in vector meson masses could be investigated directly either by
photo- or electroproduction of these mesons from nuclei, or from the decay
$\rho\rightarrow e^+e^-$ in warm, dense matter produced in relativistic
heavy-ion collisions.

\section*{Acknowledgments}
I am most grateful to M. K. Banerjee for patiently teaching me about PCAC, and
to J.~McGovern for collaboration on the topics described here and for a
critical reading of the manuscript. I am indebted to the ECT$*$, Trento for its
hospitality and stimulating environment during the workshops on Chiral Symmetry
in Hadrons and Nuclei, the Quark Structure of Baryons, and Mesons and Baryons
in Hadronic Matter. My understanding of the ideas presented here has benefitted
 from discussions with many of the participants at those workshops. In
particular I should like to thank T.~Cohen, M.~Ericson, M.~Rho, G.~Ripka,
D.~Riska, M.~Soyeur, I.~Towner and W.~Weise. This work was supported by an SERC
Advanced Fellowship.

\normalsize

\newpage

\section*{Figure captions}

\vspace{10pt}
\noindent Fig.~2.1 The Mexican hat potential for the meson fields in the
linear sigma  model.
\vspace{20pt}

\noindent Fig.~2.2 Contributions to pion-nucleon scattering in the linear sigma
model.
\vspace{20pt}

\noindent Fig.~2.3 Direct $\sigma$ exchange diagram.
\vspace{20pt}

\noindent Fig.~2.4 Diagrams corresponding to two-pion exchange in the linear
sigma model: (a) pion loop, (b) pion vertex correction, (c) crossed box, and
(d) box.
\vspace{20pt}

\noindent Fig.~3.1 The energy density of a Fermi gas of nucleons in the  linear
sigma model (3.3). The potential energy of the meson fields is shown by the
solid line, the fermionic energy by the the dashed line, and their sum by the
dot-dashed line.
\vspace{20pt}

\noindent Fig.~3.2 The density dependences of the mean sigma field for
Fermi gases of quarks (solid line) and nucleons (dashed line). These are
for a sigma mass of 600 MeV and include explicit symmetry breaking.
The density is expressed in terms of nuclear matter density, $\rho_0=0.17$
fm$^{-3}$.
\vspace{20pt}

\noindent Fig.~3.3 Contributions to pion scattering from two nucleons in the
linear sigma model: (a) rescattering, (b-d) one-pion irreducible processes.
In (a) each blob denotes the sum of the three diagrams of Fig.~2.2.
\end{document}